# Future mmVLBI Research with ALMA:
# A European vision

*R.P.J. Tilanus[23,2], T.P. Krichbaum[44], J.A. Zensus[44,51], A. Baudry[7], M. Bremer[29], H. Falcke[23,44], G. Giovannini[26,14], R. Laing[15], H. J. van Langevelde[39,2], W. Vlemmings[48]*

Contributing authors:

Z. Abraham[64], J. Afonso[10], I. Agudo[39], A. Alberdi[35], J. Alcolea[47], D. Altamirano[3], S. Asadi[13], K. Assaf[38], P. Augusto[10], A-K. Baczko[30], M. Boeck[44], T. Boller[43], M. Bondi[26], F. Boone[52], G. Bourda[7], R. Brajsa[21], J. Brand[26], S. Britzen[44], V. Bujarrabal[47], S. Cales[63], C. Casadio[35], V. Casasola[26], P. Castangia[28], J. Cernicharo[8], P. Charlot[7], L. Chemin[7], Y. Clenet[6], F. Colomer[47], F. Combes[41], J. Cordes[62], M. Coriat[65], N. Cross[53], F. D'Ammando[26], D. Dallacasa[14], J-F. Desmurs[47], R. Eatough[44], A. Eckart[51,44], D. Eisenacher[30], S. Etoka[58], M. Felix[5], R. Fender[55], M. Ferreira[9], E. Freeland[13], S. Frey[17], C. Fromm[44], L. Fuhrmann[44], K. Gabanyi[42], R. Galvan-Madrid[15], M. Giroletti[26], C. Goddi[39], J. Gomez[35], E. Gourgoulhon[6], M. Gray[38], I. di Gregorio[15], R. Greimel[22], N. Grosso[46], J. Guirado[12], K. Hada[26], A. Hanslmeier[22], C. Henkel[44], F. Herpin[7], P. Hess[19], J. Hodgson[44], D. Horns[58], E. Humphreys[15], B. Hutawarakorn Kramer[43], V. Ilyushin[34], V. Impellizzeri[69], V. Ivanov[66], M. Julião[9], M. Kadler[30], E. Kerins[54], P. Klaassen[2], K. van 't Klooster[71], E. Kording[23], M. Kozlov[50], M. Kramer[44,38], A. Kreikenbohm[30], O. Kurtanidze[1], J. Lazio[68], A. Leite[9], M. Leitzinger[22], J. Lepine[64], S. Levshakov[50], R. Lico[26], M. Lindqvist[48], E. Liuzzo[26], A. Lobanov[44], P. Lucas[59], K. Mannheim[30], J. Marcaide[12], S. Markoff[3], I. Martí-Vidal[48], C. Martins[9], N. Masetti[24], M. Massardi[26], K. Menten[44], H. Messias[10,63], S. Migliari[11], A. Mignano[26], J. Miller-Jones[67], D. Minniti[33,61], P. Molaro[49], S. Molina[35], A. Monteiro[9], L. Moscadelli[27], C. Mueller[30], A. Müller[16], S. Muller[48], F. Niederhofer[15], P. Odert[22], H. Olofsson[48], M. Orienti[26], R. Paladino[26], F. Panessa[25], Z. Paragi[39], T. Paumard[6], P. Pedrosa[9], M. Pérez-Torres[35], G. Perrin[6], M. Perucho[12], D. Porquet[46], I. Prandoni[26], S. Ransom[69], D. Reimers[20], M. Rejkuba[15], L. Rezzolla[31,45], A. Richards[38], E. Ros[44,4,12], A. Roy[44], A. Rushton[55], T. Savolainen[44], R. Schulz[30], M. Silva[9], G. Sivakoff[70], R. Soria-Ruiz[47], R. Soria[67], M. Spaans[40], R. Spencer[38], B. Stappers[38], G. Surcis[39], A. Tarchi[28], M. Temmer[22], M. Thompson[59], J. Torrelles[36], J. Truestedt[30], V. Tudose[32], T. Venturi[26], J. Verbiest[18], J. Vieira[9], P. Vielzeuf[9], F. Vincent[6], N. Wex[44], K. Wiik[57], T. Wiklind[66], J. Wilms[56], E. Zackrisson[13], H. Zechlin[60]

*(Affiliations are listed at the end of the report)*

## 1. Summary

Very long baseline interferometry at millimetre/submillimetre wavelengths (mmVLBI) offers the highest achievable spatial resolution at any wavelength in astronomy. The anticipated inclusion of ALMA as a phased array into a global VLBI network will bring unprecedented sensitivity and a transformational leap in capabilities for mmVLBI. Building on years of pioneering efforts in the US and Europe the ongoing ALMA Phasing Project (APP), a US-led international collaboration with MPIfR-led European contributions, is expected to deliver a beamformer and VLBI capability to ALMA by the end of 2014. Moreover, in Europe additional funding has become available through an ERC synergy grant (BlackHoleCam) in support of mmVLBI observations of Sgr A* and M 87.

This report focuses on the future use of mmVLBI by the international users community from a European viewpoint. Firstly, it highlights the intense science interest in Europe in future mmVLBI observations as compiled from the responses to a general call to the European community for future research projects (see section 4). A wide range of research is presented that includes, amongst others:
- Imaging the event horizon of the black hole at the centre of the Galaxy
- Testing the theory of General Relativity an/or searching for alternative theories

- Studying the origin of AGN jets and jet formation
- Cosmological evolution of galaxies and BHs, AGN feedback
- Masers in the Milky Way (in stars and star-forming regions)
- Extragalactic emission lines and astro-chemistry
- Redshifted absorption lines in distant galaxies and study of the ISM and circumnuclear gas
- Pulsars, neutron stars, X-ray binaries
- Testing cosmology
- Testing fundamental physical constants

Although a brief summary of the main research topics is included, this paper specifically does not aim at a repetition of the comprehensive science case for the ALMA Phasing Project (APP: Fish et al. 2013, arXiv:1309.3519).

Throughout this document we will use the term mmVLBI to refer to VLBI at frequencies above ~30 GHz ($\lambda \sim 10$ mm), e.g. 43 GHz ($\lambda \sim 7$ mm), 86 GHz ($\lambda \sim 3.5$ mm), 230 GHz ($\lambda \sim 1.3$ mm), and higher.

## 2. Introduction

The enormous scientific potential of VLBI at millimetre and submillimetre wavelengths is that it enables the highest resolution imaging currently possible at any wavelength in astronomy, achieving about 20 microarcsecs for a 9000 km baseline at 1.3 mm. Years of pioneering collaborative efforts and investment in hardware development were necessary, primarily in Europe (led by the group at MPIfR) and in the US (led by Haystack Observatory). They have paved the way for a matured state of the applicable techniques, where near-'routine' imaging observing campaigns with adequate detection sensitivity are now within reach. Thus, the feasibility of a long awaited common-user accessible observing opportunity is no longer a long-term guiding goal but rather a tangible and achievable mid-term goal. The ALMA Phasing project is a key development. It is expected to prepare ALMA for mmVLBI within the upcoming two years, with initial tests possibly starting in 2014. Using ALMA as a phased array to participate in a global mm VLBI network will offer unprecedented sensitivity in a largely unexplored spectral domain and at very high angular resolution. This leap in capability will enable the imaging of the shadow of a black hole, the origin of relativistic jet flows in active galactic nuclei (AGN), the innermost region of galaxies, and dusty winds near stellar surfaces. Its range of astronomical observations potentially supports addressing a wide range of fundamental questions from General Relativity in its strong-field limit to improved measurements of the Hubble constant, astrochemistry at early epochs of the universe through high angular resolution VLBI studies of extragalactic absorption lines, and the physics and dynamics of evolved stars or star-forming regions through VLBI observations of cosmic masers.

A main driver behind the APP and current 1.3-mm VLBI observations has been the "Event Horizon Telescope" (EHT), which currently is a US-led and PI-based international collaboration. Europe's prime contributors and active partners in the EHT so far have been the MPIfR and IRAM through the participation of the IRAM antennas and the APEX telescope in Chile. As a recent development, the European Research Council awarded a Synergy Grant to a EU-led and PI-based project, "BlackHoleCam", involving Radboud University, MPIfR, AEI Potsdam, Univ. Frankfurt, JIVE, MPE Garching, and ESO to support global mmVLBI and pulsar observations. It is expected that both projects will naturally complement each other and indeed join forces, and the formation of an international charter-based consortium is underway. Once commissioned, phased ALMA will be available to the global ALMA community through open competition, both for VLBI as well as a stand-alone facility. In preparing to adequately develop mmVLBI as a global facility, both in terms of implementing the necessary infrastructure within Europe, as well as its scientific use by European groups, it is now opportune and timely to organize and poll the mmVLBI community in Europe.

An initial step in this process was the ESO Workshop on "mm-wave VLBI with ALMA and Radio

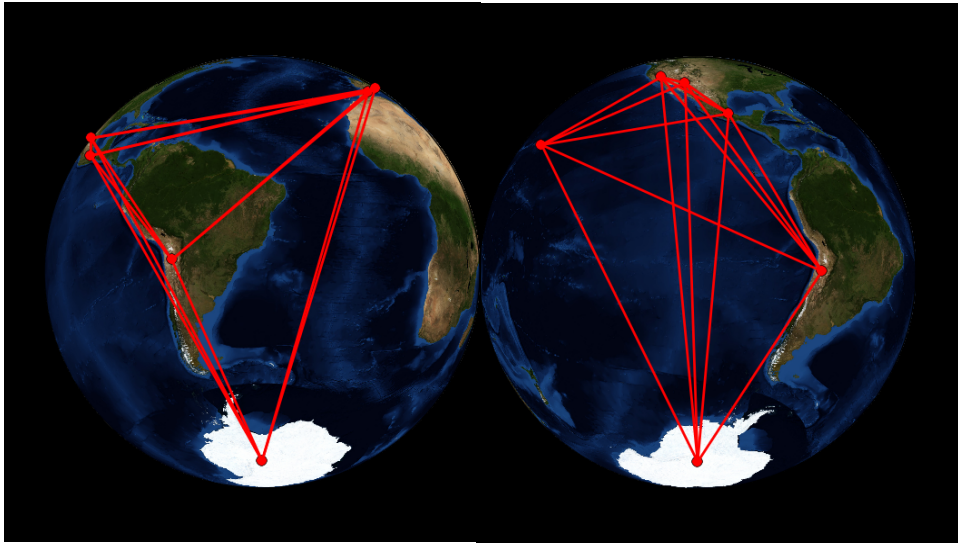

Figure 1: mmVLBI baselines at 230 GHz as viewed from Sgr A*

Telescopes around the World" (Garching, 26-27 June 2012)[1] that was attended by 61 scientists from the EU and 5 from outside Europe. The workshop highlighted the strong interest within the community in beam-formed observations with ALMA and the recognition that such a facility will represent a transformational leap in capabilities for both mmVLBI and pulsar research.

There was general agreement at the conclusion of the meeting that, from a European perspective, the ultimate goal of the planning process should be a facility accessible to the general user, with time made available through open competition. On the way towards this goal, the scientific needs of the user community need to be compiled as a necessary input for a future and formal involvement of ALMA in mmVLBI and other beam-formed science. Some of the technical challenges of this project are being addressed by the international multi-partner APP: the task for the nascent European mmVLBI user community is to organize itself to prepare for the successful use of this exciting new capability.

This document summarizes and highlights the wide range of science interests and needs of the European users, as compiled from the their responses to a general call for statements of interest in future mmVLBI. It also outlines what we see as a roadmap toward a future mmVLBI collaboration.

## 3. Summary of the general science case

Here we provide a brief summary of the general science case. For a comprehensive overview see the publication by Fish et al. 2013 (arXiv:1309.3519) and references therein.

The reason why the millimetre-wave region is so important for ultra-high-resolution imaging is that in the m/sub-mm band heterodyne receivers can still be used and quantum effects do not yet limit our ability to do interferometry, as they do at shorter wavelengths. The addition of ALMA offers a leap in sensitivity, thus making many more and weaker sources accessible to ultra-high-resolution studies and greatly improving image fidelity for the brighter objects. Only sources with flux densities above several hundred mJy have been observable so far with VLBI at 3mm; this threshold will be reduced

---

[1] Falcke, H., Laing, R., Testi, L., Zensus, A. 2012, Report on the ESO Workshop "mm-wave VLBI with ALMA and Radio Telescopes around the World", The Messenger, vol. 149, p. 50-53

by up to two orders of magnitude by a mmVLBI array that includes ALMA and exploits the larger bandwidths delivered by the current generation of mm receivers. ALMA's wide frequency range, covering all mm and sub-mm windows, will support a broad range of VLBI science to be pursued by astronomers in Europe and around the world.

Another main reason why observing in the millimetre regime is important, is that towards the Galactic Centre scattering by the interstellar medium at long radio wavelengths broadens the observed size of sources such as Sgr A* (van Langevelde et al. 1992, ApJ 396, 686). This scattering decreases at higher frequencies and vanishes at frequencies above roughly 200 GHz. Thus mmVLBI can penetrate the blurring ISM and allows a direct imaging of the underlying structures with its unsurpassed angular (and spatial) resolution.

One major science case that is driving the development of mmVLBI is the possibility of imaging the innermost parts of accretion discs and jets around supermassive black holes and, in particular, imaging the shadow of the event horizon in the black hole at the Galactic Centre, Sgr A* (Falcke et al. 2000, ApJ, 528, L13; Doeleman, et al. 2008, Nature, 455, 78). mmVLBI is the only technique available at present to directly observe nearby super-massive BHs on event-horizon size scales. Sgr A* is the most favourable source for this experiment since the angular size of its black hole shadow on the sky is the largest known and has optically thin mm and sub-mm emission on event-horizon scales. The shapes and sizes of the shadow and the expected photon ring surrounding it are precisely predicted by Einstein's theory of General Relativity (GR). Thus this observation will provide tests not only of the black hole paradigm but also of GR in its strong-field limit.

The high sensitivity of a phased ALMA will also enable the detection of short-time variability through a direct monitoring of the closure phase signal. Depending on the spin of the BH in Sgr A*, quasi-periodic variability is expected on a time scale of 5-30 minutes for compact structures rotating in the accretion disc. The sensitivity of ALMA is needed to reliably detect the small variations in these closure phases.

Another target of major interest in this respect is the super-massive black hole in the central galaxy of the Virgo cluster, M 87. This black hole is 2000 times further away than Sgr A*, but also 800 times more massive and their event horizons are comparable in angular size. Moreover, in contrast to Sgr A*, the central black hole in M 87 is accreting material at a much higher rate and is generating a pronounced relativistic plasma jet, with internal shocks that produce the observed broadband radio to TeV emission. Recent mmVLBI observations (Krichbaum et al. 2007, "Exploring the Cosmic Frontier: Astrophysical Instruments for the 21st Century", 189; Hada, et al. 2011, Nature, 477, 185) indicate that this kpc-scale jet forms on the scale of only a few Schwarzschild radii, thus imposing extreme constraints on possible models for jet formation and how the jet connects to the BH/disk system. In addition to M 87, jets formation and acceleration can be studied in nearby microquasars and also in some of the more distant AGNs where polarimetry can provide information on the detailed morphology and the role of magnetic fields in jet formation. In general, VLBI at 230 GHz and above will reveal the region where the jets are formed, thereby addressing one of the fundamental problems of astrophysics: jets and accretion discs are ubiquitous in astronomical objects ranging from young stars to supermassive black holes. However, how jets precisely form is still poorly understood. The elaborate numerical simulations available today critically need more high spatial resolution observations.

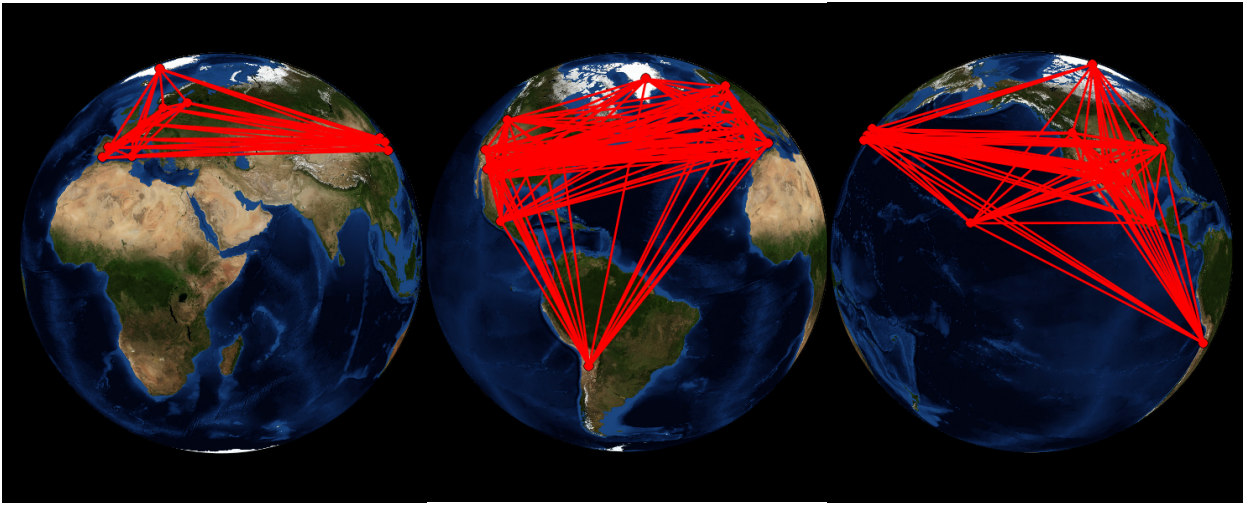

Figure 2: mmVLBI baselines at 86 GHz as viewed from M 87. The figure includes baselines with GMVA stations under development, the Sardinia Radio Telescope and the Large Millimeter Telescope in Mexico, and a future Greenland Telescope.

Another fascinating science driver at millimetre and sub-millimetre wavelengths is high-brightness, strong non-LTE line emission from a broad variety of objects (see e.g. a review by Humphreys 2007, IAU Symp. 242, 471 and references therein). Well-known examples of maser sources occur in evolved stars, where $H_2O$ and SiO masers provide impressively detailed tracers of the dynamics of the stellar envelopes or of the extended photospheres and molecular and dusty winds blown away from the old star into the interstellar medium. Some of the maser lines of water and other molecules above 35 GHz are expected to come from the most excited molecules close to the stellar surface. Parallaxes and proper motions of masers associated with stars can also be determined to derive distances and galactic motions throughout the Galaxy independently of other methods. Masers are also seen in star-forming regions and young stellar objects, particularly in regions excited by shocks. They may trace infall through accretion disks as well as outflow through jets and sensitive enough mmVLBI may allow for precise measurement of the dynamics of these processes.

Extragalactic maser sources are also known. A famous example is the nearby active galaxy NGC 4258, where a warped molecular disc surrounds the central supermassive black hole. VLBI proper motion observations of 22 GHz water masers in the disc have been used to obtain a precise geometric distance to this galaxy, making it a crucial component of the extra-galactic distance ladder. Currently the water maser cosmology project is extending this technique to larger distances, thereby providing an independent measurement of the Hubble constant and a direct determination of black hole masses. In NGC 3079, a nearby Seyfert2/Liner galaxy, and in Arp 220, a nearby megamaser, water maser emission has been detected at 183 GHz. This line, the 439 GHz line, also detected in NGC 3079, and a few other submm $H_2O$ lines may trace the inner regions of the accretion disc.

Spectral-line mmVLBI can be used to observe molecules in absorption against strong background quasars for studies of the circumnuclear matter and the intervening ISM and measurements of fundamental physical coupling constants. The sensitivity enhancement offered by ALMA will allow probing the matter and chemistry of the inner regions in galaxies and by this increase our understanding of galaxy evolution and AGN feedback.

There are many other compact sources that are potential targets for mm-wave VLBI. Of particular interest are transient sources such as supernovae, gamma-ray bursts and microquasars since the highest frequencies probe the earliest phases of any eruption and the direct environment surrounding these objects.

For stand-alone phased array observations searches for pulsars in the Galactic Centre stand out. There the low-frequency pulsed emission suffers from huge dispersion and scattering, to the extent that many pulsars become undetectable. At higher frequencies ALMA changes the prospects of finding a pulsar near the Galactic Centre completely: phased ALMA will have a sufficiently large sensitivity to perform very deep searches of the Galactic Centre even for normal steep-spectrum pulsars. An optimally located pulsar could measure the black hole mass to one part in a million, provide independent tests of GR, as well as e.g. the , "cosmic censorship" conjecture and the "no-hair" theorem, which states that all BH properties, including the quadrupole and higher-multipoles of the spacetime, are determined only by the mass and spin of the BH.

## 4. Contributed statements of research interest

Early in 2013 the editorial board of this paper issued a general call to European researchers, with emphasis on the current VLBI community, soliciting for concise statements of interest describing future interest in observations with a mmVLBI array that includes ALMA operating in its mm/submm bands. The response exceeded our expectations: about 50 projects from more than 160 researchers affiliated with 70 institutes were submitted. The proposed research spans a very wide range of topics, significantly expanding on the general science case. In collaboration with the PIs, each statement was condensed into a summary for inclusion in this report. The resulting contributed statements of interest are listed in Appendix A as C01 to C49. It is to be expected that some of these proposals may generate new multi-regional or international collaborations or may extend existing worldwide collaborations.

Concise statements rather than full proposals were solicited from the community, so detailed specifications of the capabilities of a future mmVLBI array were not provided. We therefore did not conduct a comprehensive assessment of the contributed statements for technical feasibility.

In conclusion, although we do not attempt to provide a rigorous European science case for a future mmVLBI array, the list of research topics as presented in Appendix A highlights the diverse and broad interest in such a facility within the European astronomical community. As such it emphasizes that within Europe there is ample support and a broad platform for the development of an infrastructure or facility that will give European researches access to state-of-the-art mmVLBI observations in the ALMA era.

## 5. Future mmVLBI facility and Roadmap

In this section we present the outline for a roadmap towards a future international mmVLBI collaboration[2] that includes ALMA. Such collaboration must be global in order to provide an array with the largest possible synthesized aperture and with a maximum number of participating stations. While some observatories will provide more sensitivity, it is only in combination with other distant stations that they can bring this advantage into effect. Especially at frequencies above 100 GHz, observatories become scarce and the collaboration needs to ensure to include as many as possible.

---

[2] The generic terms 'facility' and 'network' will be used as synonyms for a global mmVLBI Collaboration

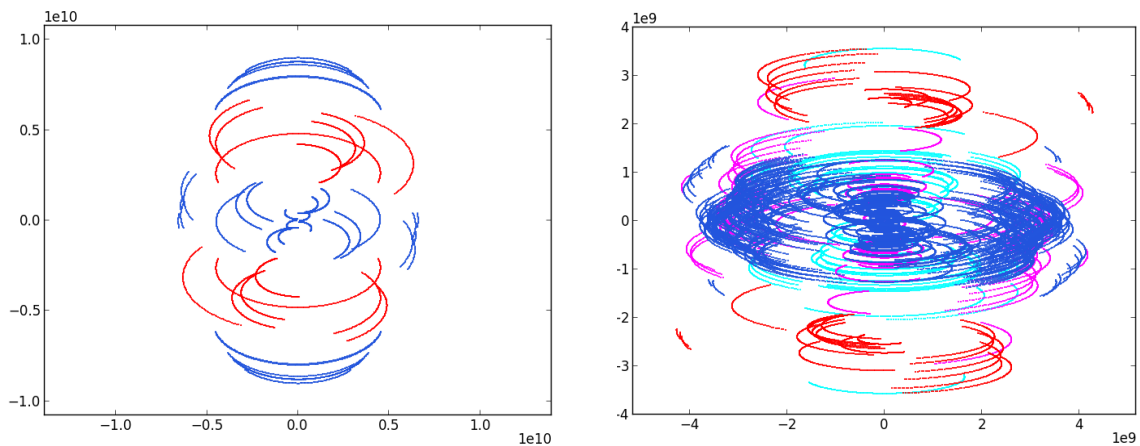

Figure 3. *Left:* UV coverage for EHT 230 GHz observations of Sgr A*; *Right:* UV coverage for 86 GHz observations of M 87. Baselines with ALMA are shown in red; Tracks with VLBA stations under development, the Sardinia Radio Telescope and the Large Millimeter Telescope in Mexico, are magenta. Tracks with a future Greenland Telescope are in light blue.

The development of the overall roadmap towards such collaboration must therefore be a multi-regional effort that includes partners outside Europe. In referring to a collaboration, we do not imply that this is to be a stand-alone operation, observatory, or VLBI network. Instead we mean a coordinating body to manage proposals and observations, giving researchers access to a network that links observatories around the world for global VLBI in joint mm and sub-mm projects. It is likely that such a global VLBI collaboration will be formed through a cooperation between existing local networks supplemented with currently still isolated (sub-)mm facilities. Later in this chapter we will develop some scenarios on how such institutes with very different sensitivities and local over-subscription factors may be combined.

In anticipation of the international process to define its operations, we also present a European vision on the aspects and principles that we propose as general guidelines for a mmVLBI collaboration. In presenting guiding principles for its operations we emphasize that they apply to the mature facility. While VLBI at centimetre wavelengths has already reached this point, as is demonstrated by well-developed facilities and networks such as e.g. the VLBA and EVN, that is not yet the case for millimetre VLBI. The GMVA offers global observations up to 94 GHz ($\lambda < 3.2$ mm), but none of the existing VLBI networks presently offers observations at higher frequencies such as 230 GHz ($\lambda \sim 1.3$ mm). VLBI at these higher frequencies involves a different set of observatories for some of whose VLBI capabilities are still in the process of being implemented. Following on to earlier single baseline pilot studies, since 2007 a collaboration that is now forming the core of the Event Horizon Telescope (EHT) has carried out dedicated 230 GHz VLBI observations using a number of ad-hoc array configurations about once a year. The EHT and other groups are also in the process of implementing instrumental upgrades that that are relevant for a future common-user network at frequencies above 100 GHz.

A key-point here is that the timeline on which a beam-formed ALMA will become available to the user community likely implies that initial mmVLBI observations will have to be conducted with a transitory infrastructure that will need to be developed into the mature long-term facility. Such a facility must also take into account that observations at higher (i.e. above 100 GHz) frequencies are increasingly weather dependent with shorter atmospheric coherence times and may require specific scheduling criteria as well as specialized hardware (wide-band backends, high-quality frequency standards). From the beginning, ALMA should take an active role in the implementation of the mmVLBI collaboration.

## 5.1 Vision

As summarized in section 3 of this paper, a comprehensive science case for future mmVLBI with ALMA was presented by Fish et al. 2013 (see arXiv:1309.3519). Section 4 highlights the broad scientific interest in Europe in future mmVLBI observations. Based on conclusions resulting from the 2012 ESO workshop on mmVLBI (Falcke et al. 2012, see footnote 1), we propose the following aspects as guiding principles for such a collaboration, bearing in mind the large user interest in Europe for its capabilities:

- To be an international collaboration, possibly with independently funded regional representation where desired.
- To offer 'open' access with time allocation based on scientific merit as judged through peer review of proposals.
- To be available to non-expert observers.
- To have a form of dynamic scheduling allowing for a flexible response to weather conditions and other operational issues.

These principles should be used as guidelines for the governance structure of the mmVLBI Collaboration, dictate the development of its infrastructure, and be part of the agreements with the participating observatories.

Section 5.2 will discuss in more detail the main implications of each of these guidelines.

## 5.2 Requirements

### 5.2.1 International collaboration

The future mmVLBI Collaboration will provide the global framework for (sub-)mm observatories from around the world to come together in a VLBI network for astronomical observations in the frequency range spanned by the ALMA observing bands. At the low end of the range, i.e. ALMA band-1 at 35 to 52 GHz ($\lambda \sim 7$ mm), an observation could involve 10-20 participating facilities providing an excellent imaging capability and excellent atmospheric transparency at all stations. The number of available stations drops quite steeply towards the higher frequencies and at frequencies of 350 GHz ($\lambda \sim 0.8$ mm) or even higher observations with limited imaging capability may be possible. The mmVLBI Collaboration will handle the logistical complexities of the network and make mmVLBI observations available to the international astronomical community. A summary of its main functions is:

- Setup and administer a Memorandum of Understanding/Agreement (MoU/MoA) with each of the participating observatories and processing centres.
- In collaboration with the participating facilities develop, implement, and maintain the infrastructure of the mmVLBI network.
- Arrange regular VLBI sessions with the participating facilities.
- Issue regular calls for proposals and manage the proposal submission, proposal review and time allocation process.
- Assist PIs with the set up of approved observing projects.
- Allocate and distribute VLBI recording media (disks)
- Carry out the approved program of mmVLBI projects in consultation with the PIs.
- Correlate the resulting data and carry out basic quality control in the corresponding facilities.
- Distribute the correlated and pre-calibrated data sets to the PIs.
- Archive the observations for public access after the proprietary period expires.
- Maintain performance and operations statistics.

Funding will need to be identified to resource the operations of a future mmVLBI Collaboration. An operational model similar to that of ALMA could be promoted: various aspects of the operations could be handled by local support facilities, likely associated with the various correlation centres, or regional institutes such as ESO, NRAO, and NOAJ, on behalf of their regional community.

*5.2.2 Open access with peer-reviewed proposals*

This principle is a common feature of non-proprietary astronomical observatories. The default process involves a call for proposals, submission deadline, proposal review, and a time allocation decision. The requirement of open access and peer review calls for the adoption of a policy and procedure to review and rank proposals. Traditionally, facilities participating in VLBI observations had different deadlines, different formats and requirements for proposals, and differ in other details of this process. The existing VLBI networks have successfully overcome these complications via different means, using two different operational models:

   a. Agree between the facilities on the amount of time to be reserved for VLBI during the upcoming cycle/year and allocate observing time based on a single, combined peer-review process according to own proposal deadlines. The main advantage of this option is that it bypasses the majority of the aforementioned complications by closely emulating a single observatory. The individual observatories have input in the scientific ranking of proposals through representation on a shared time allocation committee, but are committed to the consensus result. The EVN, VLBA, and Australian LBA time allocation processes operate largely in this mode.
   b. Proposals are centrally collected and forwarded to the participating observatories (we will ignore the option of PIs submitting to each observatory themselves). Each observatory decides how it wants to referee VLBI proposals, be it through their own or a shared process. A proposal's acceptance will depend on the combined ranking across the network. Individual observatories can opt to be excluded from specific projects. The advantage of this model is that it leaves individual flexibility and accountability to each of the participating observatories. A main disadvantage is an exposure to most of the aforementioned complications and this mode is only practical if the proposal cycles and refereeing processes at each of the observatories follow a similar time-line. Also, if a single large observatory, such as ALMA, vetoes a project it may well become unfeasible. The GMVA presently operates along these lines. The High Sensitivity Array (HSA) and the Global array (EVN+VLBA) operate in a mixed form.

These two options represent the extremes. A variation that leaves more flexibility for the individual observatories, but is intermediate between the two options above, is:

   c. As outlined in option `a', but give individual observatories the option not to participate in the observations for a specific proposal in case an observatory's assessment of the proposal ranking significantly differs from that of the combined review. Obviously, such decision should be a well-justified exception, but will leave observatories with a measure of independent control.

Other checks and balances can be added, but likely the time allocation process of a future mmVLBI Collaboration will fall in between the two extreme options outlined.

Regardless of what option is adopted, the major millimetre observatories will remain in strong demand and likely will see high over-subscription rates for the foreseeable future. Consequently, the amount of time available for mmVLBI can be expected also to remain limited initially. This suggests to start with a relatively simple, low-overhead process for proposal submission and time allocation with a provision to review at set intervals and to adapt the model to reflect a maturing mmVLBI facility.

*5.2.3 Available to non-expert observers*

This aspect sets a minimum requirement on the infrastructure of the mmVLBI Collaboration, as it requires transparent and streamlined access by any user to the observation details and data. In practice this implies that VLBI observations:

- Will need to be carried out in a service mode that does not rely on the presence of the proposer.
- Are processed by the correlations centres with adequate quality control, in the initial fringe search and especially in the calibration process
- Are delivered to the proposer in a common, well-documented data standard (i.e. FITS-IDI).
- Are supported by adequate assistance to guide less experienced proposers and observers.

The European ARC node model might be a good template for setting up a network of regional support for mmVLBI that leverages available resources that are distributed geographically.

*5.2.4 Triggered scheduling*

A future mmVLBI network will not be a dedicated array that can relatively straightforwardly shift projects in response to observing conditions. Nevertheless, some form of dynamic scheduling will still be important given that atmospheric conditions have a major impact on the observability at mm-wavelengths. This is especially the case for observations above 200 GHz, where the chance for suitable weather across most of the VLBI network will be poor on any random day (and the demand for the good weather by non-VLBI modes will continue influencing the policy of individual observatories). Rather than full dynamic scheduling with a relatively short-term response trigger, the go-ahead for a VLBI station will typically come well ahead of the start of observing given that the array will be deployed from east to west as the target rises and sets at the participating stations. The ability for a station to drop out of VLBI mode quickly and `dynamically' switch back to the previous mode is an essential requirement though to avoid excessive loss of observing time.

Participation in mmVLBI observing can be accommodated through the acceptance of outside triggers that will override the default telescope schedule at some point within the next 24 hours. This would be somewhat similar to the target-of-opportunity trigger in place at many observatories. For dynamically scheduled facilities this should not be an issue, although individual stations may have additional constraints as to when a trigger can be accepted. Such additional constraints will be part of the overall decision matrix to trigger mmVLBI. Overrides are more problematic for participating facilities that host visiting observers scheduled to carry out a particular program and each will have to implement a policy that formulates an acceptable compromise.

Although in principle it is possible to keep VLBI available on a continuous basis, a more common approach is to assign blocks of time during which VLBI will be on stand-by and triggered when a window arises when conditions across the array are appropriate. The decision to trigger a VLBI run will require accurate weather forecasts and status information from each of the stations. While global satellite maps are available as a rough guideline, local or even regional forecasts are much less available, especially for stations where the local topography has a large impact (e.g., mountain tops). Given the relative superior characteristics of the ALMA site, the deciding element often will be the weather forecast for critical stations at less favourable locations. Typically suitable weather comes in windows of a few days, separated by, possibly long, periods of inclement weather. The observing blocks during which VLBI is on stand-by will need to be long enough that a few of such windows can be expected to occur at each station and for a reasonable chance that there will be an overlap during the block. This suggests individual blocks of order 1-4 weeks several times per year.

Typically there will be significant lead-time before VLBI observing. Logistical considerations associated with blocks that are several weeks long dictate that the VLBI equipment will have to be turnkey in a sense that the switchover should not require the presence of 'VLBI experts'. In addition, for observatories that operate unattended for long periods, the switchover will need to be remotely controllable. This will require further development of the VLBI equipment and observing modes currently implemented at a typical sub-mm observatory.

In summary, we note that expectation of a relatively moderate number of observing nights per year in combination with its sensitivity to weather conditions places mmVLBI in a somewhat unique position. It suggests the implementation of an infrastructure that can be on stand-by over prolonged windows of several weeks and, when conditions are right, can dynamically become active when the right observing conditions arise. While the response to the trigger does not have to be automatic, the switchover has to be relatively turnkey and able to be carried out by the regular staff at each observatory. Developing this infrastructure and ability will form the core of the operational implementation of a mature mmVLBI facility.

**5.3 Existing operational networks**

The development of an international mmVLBI network builds on experience mostly gained at longer wavelengths. There are several astronomical VLBI networks in operation at present. The European VLBI Network (EVN), US-based Very Long Baseline Array (VLBA), and the Australian Long Baseline Array (LBA) are predominantly cm-wave networks, but the first two offer observations up to 43 and 86 GHz respectively with a limited number of stations. 'Global VLBI' combines the EVN and the VLBA at frequencies up to 45 GHz. The high-sensitivity array (HSA) combines the VLBA with the largest telescopes (Effelsberg, GBT, VLA, Arecibo) for Global VLBI observations with best sensitivity. The Global mmVLBI Array (GMVA), combines European (EVN and non-EVN) observatories and the VLBA and GBT and operates predominantly at 86 GHz ($\lambda \sim 3.5$ mm).

In addition, a number of other VLBI networks are in operation in Japan (JVN, VERA), Korea (KVN), and China (CNV). Developments are underway to combine these networks into the East Asia VLBI Network (EAVN), which plans to operate up to millimetre wavelengths with a subset of the stations. Although we do not discuss the LBA or the EAVN in detail, the assumption is that they will take part in the future mmVLBI array and participate in its development.

In the context of future mmVLBI with ALMA, the Event Horizon Telescope (EHT) has an important role in that it carries out pilot VLBI observations at 230 GHz ($\lambda \sim 1.3$ mm), beyond what is being offered by the aforementioned networks. The three networks and the EHT experiment will be discussed in more details below.

A number of telescopes currently under development or construction are expected to join mmVLBI observations once fully operational. Among those are the Large Millimeter Telescope (LMT), the Greenland Telescope (GLT), the Long Latin American Millimeter Array (LLAMA), the expanded Plateau de Bure interferometer (NOEMA), the South Pole Telescope and others.

We point out that many of the submm-wave antennas are located in the Americas, providing limited East-West baselines. An additional submm telescope located in Africa would solve this problem and provide an excellent sub-array that involves the European (IRAM) telescopes, ALMA and LMT. Such an effort could be considered complementary to the emerging African VLBI network (AVN) and benefit from the focus on EU-African cooperation in radio astronomy as called for by the written declaration on "Science Capacity Building in Africa: promoting European-African radio astronomy partnerships" (*P7_DCL(2011)0045*) of the EU parliament from 15-03-2012.

*5.3.1 EVN*

The EVN is a collaboration of the major radio astronomical institutes in Europe, Asia and South Africa. It is a large-scale astronomical facility that is open to all astronomers. The EVN has expanded since its inception, and now includes a total of 14 major institutes collectively operating 18 major telescopes. The data sets are correlated at the Joint Institute for VLBI in Europe (JIVE), and upon request at the MPIfR Bonn correlator. It has pioneered regular real-time and eVLBI operations. The EVN Consortium Board of Directors sets the overall policy of the EVN and the Technical and Operations Group consider issues related to technical aspects of EVN operations. Observing proposals are submitted to the EVN Program Committee for peer review. Observing time is awarded based on scientific merit and technical feasibility. The EVN observes yearly during three VLBI sessions, each 3-4 weeks long. The EVN predominantly operates in the frequency range from 1.6 – 22 GHz ($\lambda \sim 18 - 1.3$ cm), but also offers VLBI over an extended range from 0.3 - 43 GHz ($\lambda \sim 90 – 0.7$ cm) with a limited number of stations. Extensive tools for actual and prospective observers are available online. EU-sponsored personal EVN user support is available from JIVE as well as travel support to JIVE and the EVN observatories.

*5.3.2 VLBA*

The VLBA is a dedicated VLBI array operated by National Radio Astronomy Observatory (NRAO) in the United States. NRAO telescopes are open to all astronomers. The VLBA is a system of 10 identical antennas located across the US from the Virgin Islands to Hawaii, controlled remotely from the Array Operations Center in Socorro, New Mexico. Observing time on additional high-sensitivity telescopes (NRAO's phased VLA and GBT; and the 100-m Effelsberg telescope) can be requested and granted as part of the network (the HSA). The recorded data are sent from the individual stations to the correlator in Socorro. As for all NRAO facilities, proposals are peer reviewed by one of eight Science Review Panels, based on its relevant sub-discipline, with the aim to select the proposals that potentially are most valuable for the advancement of scientific knowledge. Time on the VLBA and other NRAO instruments is scheduled on a semester basis. The VLBA observes at 11 main wavelength bands from 0.3 – 96 GHz ($\lambda \sim 90 – 0.3$ cm). Online tools are available and a scientific contact person is assigned in support of each approved project. For each observing program scheduled on an NRAO telescope, reimbursement may be requested to travel to the NRAO for observations or data reduction. Recently funding for the VLBA by the NSF has come under pressure. For now continued operation has been assured. In the long term funding from sources with less open access policies, reduction in services, or even closure may be a consequence.

*5.3.3 GMVA*

The GMVA has been set up by a group of radio observatories interested in performing astronomical VLBI observations at millimetre wavelengths. It presently consists of 6 observatories in Europe and the VLBA in the USA with the 8 of its 10 antennas outfitted for 3 mm, as well as the GBT for a restricted amount of time. Data are correlated at MPIfR's correlator center in Bonn and provided to the PIs in FITS-IDI and MK4 format. The GMVA offers two prearranged observing blocks a year of up to 5 days each and operates mainly at 86 GHz ($\lambda \sim 3.5$ mm). Proposals are submitted through the NRAO proposal tool (PST) and are forwarded in parallel to the European observatories for their internal refereeing. The GMVA scheduling committee (SC), consisting of a European and a VLBA scheduler, allocates observing time for each proposal on the basis of the combined ratings received from the observatories.

*5.3.4 The EHT*

The frequency of 100 GHz is a transition point: many telescopes that operate up to this frequency are at sites that do not support higher frequency observations. Conversely, a significant number of the telescopes that operate at higher frequencies are specialized and do not deploy receivers at or below 100 GHz. In practice this means that an array that operates at higher frequencies, including the one that is used by the EHT for its 230 GHz observations, has only a limited number of stations that overlap with existing networks. The EHT pursues a science objective that is focused on the imaging of the event horizons of i.e. Sgr A* and M 87. On the other hand, in accordance with a possible role as a pathfinder project for 230 GHz observations with ALMA, it is being discussed to expand it into a global project (EHT Experiment) with a large international team and a more formal structure with the formation of oversight, science, and technical working groups. It is in the process of establishing formal MoUs with participating observatories and is acting in a technologically enabling capacity through the installation and operation of dedicated VLBI equipment. The EHT experiment uses the Haystack and Bonn correlator for its data reduction and members of the EHT team are part of the ALMA Phasing Project.

**5.4 International coordination**

The previous section discussed some of the existing VLBI networks, each observing below 100 GHz, and the EHT operating at 230 GHz. A natural question is whether one of these would be an obvious candidate to form the basis of a future mmVLBI Collaboration that will also include the EAVN and additional sub-mm observatories. For a number of reasons the answer to that question appears to be that as a single organization none of the above is in an ideal position to act in that role. One aspect is that existing centimetre networks are predominantly regionally organized, which in the context of ALMA introduces complications in identifying one to take the lead in the implementation of a future mmVLBI Collaboration. On the other hand, these networks have proven to be capable partners in cooperating for global VLBI, e.g. as the global High Sensitivity Array or HSA. The GMVA, specifically set up for global VLBI at millimetre frequencies, presently has a more limited infrastructure compared to its centimetre counterparts, owing to the fact that it operates only in two frequency band (43 and 86 GHz). The EHT, of course, is not aiming to be a general-purpose, common-user facility and has not implemented much of the infrastructure associated with such a facility.

Irrespective of these complications, none are a fundamental obstacle to leveraging the existing networks in the development of future global mmVLBI with ALMA. Indeed, it is our conclusion and suggestion that the current networks in cooperation need to formalize the formation of a global mmVLBI array, in consultation with their current and prospective constituent observatories. In effect, this might entail the evolution towards a next-generation of the GMVA that is better integrated with, and leverages infrastructure in place for, centimetre VLBI, and one that will increasingly collaborate with the EAVN and could be extended to higher frequencies and include facilities that are currently participating in the EHT.

An alternative is to implement a future mmVLBI Collaboration 'from scratch' as a new network that essentially will need to duplicate the centimetre networks at millimetre wavelength. We do not believe that this is a realistic option given the current political and economic climate that already is threatening to have significant consequences for the operations of a number of ground-based astronomical facilities. Against this background, the implementation of new capabilities through cooperation, consolidation, and utilization of existing infrastructure is the more realistic option and likely to be more effective and successful.

In summary, we propose that future mmVLBI with ALMA is to be organized through international coordination as a formal cooperation of existing VLBI networks with telescopes currently partaking in the GMVA and EHT and expanded to include future mm/sub-mm observatories.

**5.5 Roadmap towards an open mmVLBI Collaboration**

The discussion in the previous section effectively presents a formal collaboration, leveraging, and extension of the main existing VLBI networks as the top-level outline for the roadmap towards a global mmVLBI Collaboration.

As noted previously, initial mmVLBI with ALMA will have to be conducted with a transitory infrastructure that will need to be developed towards the mature long-term facility as described in the previous sections. Linking technical developments with the establishment of a global mmVLBI Collaboration, the former North American ALMA Project Manager suggested an approach that includes early science observations with a beamformed ALMA at 90 and 230 GHz and colloquially is referred to as the "McKinnon plan":

- APP & Early science:
    I. Completion of the ALMA Phasing Project (APP). The APP passed its Critical Design Review in May of 2013 and completion is expected by Q4 of 2014.
    II. Early science: beamformed ALMA and VLBI at 90 GHz[3] ($\lambda \sim 3$ mm); EHT experiment as a pathfinder project for VLBI at 230 GHz ($\lambda \sim 1.3$ mm).
- Global mmVLBI Collaboration:
    III. Develop a roadmap for the implementation of a global mmVLBI Collaboration
    IV. Establish the global mmVLBI Collaboration

This approach recognizes that the two main aspects of the overall implementation are related initiatives that each require a phased approach. On the one hand the technical commissioning of mmVLBI with ALMA, on the other the realization of a global, open mmVLBI Collaboration. In a general sense we endorse this approach and agree wholeheartedly that mmVLBI proposals should be expected to compete for time against regular ALMA proposals. We note, however, that the schedule for the APP is not well aligned with the ALMA proposal cycles. For that reason, there may be a strong case to apply for mmVLBI time on ALMA between proposal calls, as we discuss briefly below, However, we stress that a speedy formation of a global EHT consortium that adequately involves the relevant international, and in particular European parties, is a necessary prerequisite for such an application.

Also, early science observations should not be impeded by the state of development of the formal international collaboration. The loss of a few stations, either due to retirement or budget issues, will directly impact the imaging capability of the array. This can be the case for the proposed 230 GHz ($\lambda \sim 1.3$ mm) pathfinder observations of Sgr A* by the EHT experiment. The direct imaging of a black hole will be an important scientific breakthrough that will undoubtedly generate widespread professional interest as well as capture the attention of the general public. But also at 86 GHz ($\lambda \sim 3$ mm) the loss of, e.g., the VLBA would have serious repercussions for the scientific capabilities of the mmVLBI array, which only in part could be compensated by a combined European GMVA and EAVN.

In order to enable a start of mmVLBI science observations independently of the facility development, it will be a necessary part of the roadmap exercise to design a robust interim framework for operations. Potentially, although possibly optimistically, ALMA may technically be able to do mmVLBI in 2015, which means that the broader aspects of such framework will need to be agreed upon on relatively short notice to be available for review by the relevant governing bodies of participating facilities. In addition, there is an urgent need for a framework within which preparations for early mmVLBI observations can proceed.

In this context, a team led by MPIfR and composed of the European GMVA institutes and several ARC nodes in Europe is carrying out an ESO upgrades development study for ALMA, with the aim of defining the joint operations between the GMVA and the phased ALMA at 3mm (and potentially at 7mm). Considering the overall plan, below we formulate an initial proposal for a default framework for initial mmVLBI with ALMA. This proposal is intended as a first step in a fast-tracked global process of review and iteration. Once an initial version is in place, the framework can evolve along the lines set out by a subsequent roadmap or eventually be replaced by a more advanced version.

The framework we propose aims at striking a balance between the timescales available, leveraging and building upon existing infrastructure, management and time-allocation overheads, and the amount of observing time that realistically can be expected for this initial phase of mmVLBI. It does not aim to implement all aspects of a mature facility, but instead to put an infrastructure in place that will be capable of supporting early mmVLBI observations in manner that conforms to the McKinnon plan (with the assumption that the EAVN at some point will become a future participant of mmVLBI as well). The key-points are:

- For early science observations at 86 GHz, add ALMA to the existing GMVA network while enhancing the proposal submission and user support aspects through utilization of the VLBA and EVN infrastructures. ALMA would then issue a call for early science proposals to its general user community. A study proposal to define this process in detail has been accepted by ESO.

---

[3] As stated, Phase II reflects a more current formulation of the plan that aims to give the user community access to beamformed science and mmVLBI at 90 GHz relatively early on.

The correlations centres at MPIfR and Socorro, involved in current 3-mm VLBI, are suggested as the default option for the data reduction. Once the band-1 ALMA receivers have been commissioned, 43 GHz VLBI can also be added.

- Add ALMA to the EHT for 230 GHz pathfinder observations (EHT experiment). This is likely to greatly shorten the timescale on which 230 GHz can be made available to the general community since it capitalizes on expertise and infrastructure the EHT has been building up throughout the past decade. The EHT experiment consortium will need to submit an observing proposal and compete for time like a regular project. Once it becomes available for general observing, 230 GHz VLBI can be offered and handled the same as 86 GHz VLBI, effectively expanding the frequency range of the GMVA-centred framework or its successor.

In effect, the proposal identifies the GMVA and EHT as the default VLBI arrays to which ALMA will be added during the start-up phase and while the development of a new international collaboration is ongoing. In their current form, neither the GMVA nor the EHT experiment, however, is fully suited for that role and will need to address a number of issues.

The main issue that the GMVA should address is user support, through synergistic use of the existing infrastructures, based on the experience of the Bonn correlator and its established partners (Haystack, the VLBA, and JIVE). In this respect we note the recent adoption of the VLBA proposal submission tool for GMVA proposals.

For frequencies over 100 GHz, we suggest an active participation of ALMA, in close collaboration with the EHT, in the formulation of pilot observations that act as a pathfinder to further develop ALMA's mmVLBI capability at e.g. 230 GHz. As mentioned, completion of the formation of a global EHT consortium is a necessary prerequisite.

Regarding early science observations we want to highlight a potentially serious complication: by default, ALMA operates an annual proposal cycle, but Cycles 1 and 2 will be substantially longer than 1 year. The most recent proposal deadline (Cycle 2, in December 2013) was for observations between June 2014 and October 2015. ALMA could in principle be ready for mmVLBI observations in Q1-Q3 of 2015 (ALMA Cycle 2), as e.g., stated in the presentation of the McKinnon plan. There is a requirement that modes offered in a given Cycle should have been tested adequately before the Call is issued, so it is by no means clear that beamformed science as well as early science 90 GHz and 230 GHz pathfinder mmVLBI could realistically be included even in the call for Cycle 3. We believe that there are strong scientific and operational reasons (e.g. the availability of other stations) to propose for ALMA time outside of the normal process. We therefore ask the ALMA Project to look, if necessary, for a suitable alternative approach that will allow its user community to exploit these exciting new capabilities of ALMA at the earliest possible opportunity. An application for Director's Discretionary Time is one such mechanism.

**5.5.1 European infrastructure developments**

Section 5.3 discusses the main existing VLBI networks in Europe: the EVN, with JIVE as its main correlation center in Dwingeloo, and the GMVA using the correlator at MPIfR in Bonn. The two facilities fulfil two essential and complementary roles. The MPIfR with collected experience over four decades has been at the forefront of the development of global mmVLBI interferometry. This exploratory role of MPIfR activities is expected to continue over the next several years, reflected in a significant participation in the ALMA Phasing Project and in the contribution of one of two places where data from the EHT and possibly the EHT experiment are being correlated. As outlined in the previous section, the GMVA is expected to play an important part during the development of a common-user mmVLBI Collaboration.

The EVN, on the other hand, operates as a mature open-user facility for centimetre VLBI and JIVE is currently transitioning towards the status of official recognition as a European research infrastructure provider. As such, JIVE will remain the main European correlator center for the foreseeable future and it is to be expected that it will become an integral component of a common-user mmVLBI Collaboration. A possible and attractive scenario is to share infrastructure and expertise by JIVE and the MPIfR at mm-wavelengths, continuing and expanding present GMVA activities.

Adapting mmVLBI at frequencies above 100 GHz ($\lambda < 3$ mm) to "ALMA" standards will require technical developments and the procurement of equipment. Examples are the new Mark 6 recorders capable of recording 16 Gbps, backends capable of sampling of 4-GHz wide bands as well as possibly hydrogen maser clocks and in-beam water vapour radiometers. In addition software developments will be needed in support of the new technical capabilities. Significant funding for such efforts is provided by the BlackHoleCam project.

The various developments will need to draw on resources that are distributed within Europe and that will need to be coordinated with their counterparts on the global stage. We suggest the formation of a steering committee to oversee and manage these development activities within Europe. Such committee could potentially be formed as a networking activity within Radionet or its successor. It could also act to represent Europe for the development of the global mmVLBI Collaboration.

In summary, we propose the formation of a European mmVLBI steering committee that will:

- Act as a coordinating body for mmVLBI developments within Europe.
- Act as a representative body for European mmVLBI on the global stage.
- Formulate and oversee, on behalf of the European user community, an integrated approach towards future mmVLBI that involves, within Europe, GMVA/MPIfR, EVN/JIVE, and other current VLBI facilities.

**APPENDIX A: Contributed statements of interest in future mmVLBI research**

The summaries below are based on contributions received in response to a general call to European researchers, with emphasis to the current VLBI community, soliciting for concise statements of interest describing future interest in observations with a mmVLBI array that includes ALMA operating in ALMA bands. In collaboration with the PIs, each statement was condensed into a summary for inclusion in this report. The resulting contributed statements of interest are listed in Appendix A as C01 to C49. Not having provided detailed specifications of the capabilities of a future mmVLBI array we did not conduct a comprehensive assessment of the contributed statements for technical feasibility.

*C01: mmVLBI and GRAVITY*
Y. Clenet[6], T. Paumard[6], G. Perrin[6], E. Gourgoulhon[6], F. Vincent[6]

Millimetre-wavelength event-horizon imaging of Sgr A* and other compact objects is a fundamental goal. It is complementary with the goals of the GRAVITY project at infrared wavelengths. Our involvement in GRAVITY will inevitably lead us to use mmVLBI data when available. Furthermore, Observatoire de Paris is also very interested in using future mmVLBI data to put to the test alternative compact objects metrics as well as metrics of alternative theories of gravitation. Comparing VLBI data to simulated data computed by the Observatoire de Paris ray-tracing code GYOTO could perform such tests.

*C02: Event Horizon Imaging*
H. Falcke[23,44], R. Tilanus[23,2] and mmVLBI group Radboud University

Possibly the most prominent science goal for global mmVLBI observations is to obtain event-horizon scale images of Sgr A* and successful 1.3 mm VLBI observations since 2007 have made it clear that obtaining such observations within reach of a global mmVLBI array. Detailed imaging and time-domain mmVLBI observations of Sgr A* and M 87 will provide tests not only of the black hole paradigm, GR in its strong-field limit, but also of e.g. the "cosmic censorship" conjecture and the "no-hair" theorem. The central black hole in M 87, while slightly smaller, is actively accreting material and the source of kilo-parsec scale jets allowing for observations of jet launching on Schwarzschild-radius scales.

*C03: X-ray flaring and mmVLBI in Sgr A\**
N. Grosso[46], D. Porquet[46]

We have been involved for a decade in the multi-wavelength studies of Sgr A*, and in particular in the study of the flaring activity of Sgr A* in X-rays. Therefore, we are very interested in a future sub-mm VLBI facility that includes ALMA to look for time-variable structures in the accretion flow. Simultaneous observations in X-rays will allow us to constrain the location of the high-energy flares from Sgr A*.

*C04: Stability of fundamental constant*
A. Mignano[26], C. Martins[9], F. Boone[52], F. Combes[41], C. Henkel[44], V. Ilyushin[34], M. Kozlov[50], S. Levshakov[50], P. Molaro[49], D. Reimers[20]

The violation of Local Position Invariance (LPI), one of the pillars of Einstein Equivalence Principle, is predicted within the framework of grand unification theories, multidimensional theories, and scalar field theories that consider the fundamental physical constants as dynamical variables. Radio astronomy plays a crucial role in these studies because it provides the highest accuracy measurements of atomic and molecular transitions that are sensitive to changes in the fine-structure constant and electron-to-proton mass ratio. Some of the best targets are seen in absorption in front of remote quasars, since then very narrow sub-components of molecular transitions can be detected at slightly different radial velocities. mmVLBI will help to unveil the very small structure of the background continuum sources and the foreground absorbing clouds, and thus to test whether different molecules are co-spatially distributed.

*C05: Alternatives theories of Gravity*
T. Boller[43], A. Müller[16], P. Hess[19]

Einstein's General Relativity (GR) invented 100 years ago successfully describes gravitation. An algebraic extension of GR to pseudo-complex (pc) variables has been proposed, called pseudo-complex General Relativity (pc-GR). One of the important consequences of pc-GR theory is the presence of a field with repulsive properties. This has the effect that for very large masses the gravitational collapse is stopped and something what we call "gray star" is formed instead of a black hole. When applied to the Friedmann-Lemaitre-Robertson-Walker model of the Universe, pc-GR results in a finite or vanishing acceleration of the Universe for very large times. We are currently performing ray-tracing simulations based on the pc-GR theory for Sgr A* and M 87 for different accretion scenarios and viewing angles for the observer.

*C06: Revealing the polarized fine structure of AGNs with mm- and Space-VLBI*
A. Alberdi[35], M. Pérez-Torres[35], T. Krichbaum[44], E. Ros[44,4,12], J. Marcaide[12], J. Guirado[12], I. Martí-Vidal[48]

Our aim is to use mmVLBI to image the polarized fine structure of AGNs. Space-VLBI at 22 GHz and mmVLBI are complementary in their scientific goals: both are sampling the same physical regions (close to the base of the relativistic jets) but at different frequencies and opacities, characterizing the spectral behavior and the polarization signature of the different emitting regions of the sub-parsec AGN radio jets. Polarimetry with mmVLBI is a unique tool to image the intrinsic B-field configuration, helping to discriminate whether the jets are launched by MHD effects. The improved sensitivity of mmVLBI with the phased-ALMA will enable a comparison of the 86 GHz polarimetric VLBI observations with RADIOASTRON observations at 22 GHz with matching angular resolutions better than 50 microarcseconds to trace the polarized fine structure of the compact sources. Recently, it has been suggested that the 86 GHz emitting region is complex, with a three-dimensional structural. The 3D-field will be traced with the high-resolution images of the polarized structure (magnetic field in
the plane of the sky) and the determination of the Rotation Measure through the comparison of the space-VLBI and mmVLBI images (magnetic field in the direction of the LOS).

*C07: GAIA and mmVLBI*
K. Gabanyi[42], S. Frey[17]

ESA's Gaia mission will provide astrometric and photometric observations of a large sample of QSOs down to V=20 mag and is expected to provide information about the optical emission within 1 mas of the central engine, thus at a moderate redshift of ~1, within a region of 8 pc of the SMBH. Optical observations of QSOs have shown photocenter jittering of a few tens of parsecs in three sources accompanied by optical magnitude variations. Either blobs or shocks traveling in the jet or binary supermassive black holes can explain these findings. With the inclusion of the phased ALMA in mmVLBI the region where the optical emission measured by Gaia may originate can be mapped with high fidelity in the radio regime. Conservative estimates indicate that a couple of hundreds of sources will be observable to investigate whether the observed optical characteristics can be connected to features in the jet launching region revealed in the mm regime.

*C08: Celestial reference frame and non-standard AGN motions*
K. Gabanyi[42], S. Frey[17]

Selected compact extragalactic radio sources serve as fiducial points to define the celestial reference frame, but astrometric VLBI shows apparent proper motions of ~0.01-0.1 mas/yr for many. It is crucial to understand the origin of these proper motions, to define a more accurate reference frame, and to accurately study phenomena like the secular aberration drift caused by the acceleration of the solar system barycenter due to the rotation around the Galactic center. While the characteristic direction of VLBI-measured AGN proper motions seems generally connected with the ~1-10-mas scale jet structure, there are cases of significant misalignment. In particular, the proper motion direction for OJ 287 is nearly orthogonal to its radio jet seen with cm-wavelength VLBI. The relationship between the small apparent proper motions and the brightness structure could be established with sensitive mmVLBI imaging which probes the right angular scales, well within 1 mas from the central engine.

*C09: High angular resolution observations of LLAGN*
M. Giroletti[26], F. Panessa[25]

Sgr A* is the nearest LLAGN. Besides providing a great chance to reveal the shadow of the SMBH, it is an ideal case to explore the region of creation of jet/outflows and the jet-disk coupling in a low-efficiency accretion regime. High angular resolution and sensitivity in the mm-wavelength regime are needed to disentangle different components and constrain physical mechanisms at work. With phased ALMA and mmVLBI, a few more LLAGN in the local universe could also be explored, where cm-wavelength VLBI has already provided interesting clues.

*C10: mmVLBI of radio cores in blazars*
J. Gomez[35], I. Agudo[39], C. Casadio[35], S. Molina[35]

Our research group is excited about the idea of future mmVLBI including the phased ALMA. Our main collaborators in this research include Alan Marscher & Svetlana Jorstad from Boston University, and the VLBI Group of the MPIfR. Ongoing topics of research are:
- The nature of the VLBI-core and in particular to test the hypothesis that it can be identified with a recollimation shock. We have currently a running program on the VLBA for phase-reference observations of 6 blazars at multiple frequencies, including 7mm and 3mm.
- High-resolution imaging. We participate in a new GMVA proposal led by Alan Marscher aimed to study the innermost regions of some of the most energetic gamma-ray blazars.
- We participate in the coordinated multi-wavelength blazar monitoring program led by the Boston University to conduct monthly 7mm polarimetric monitoring of 36 gamma-ray blazars and is aimed to identify the high-energy emission sites and mechanisms. Our group is particularly interested in this kind of studies with the future global mmVLBI, especially if the phased ALMA is included in the array.
- mmVLBI observations of double-sided jets can provide a direct determination of the location of the central engine (black hole) relative to the jet and counter-jet footprints. We recently submitted a VLBA proposal with observations at 7mm aimed to measure the distance between the central black hole and the radio core in several sources and anticipate similar studies with a future mmVLBI array including the phased ALMA.
- We are also particularly interested in polarization studies of the magnetic field structure with the highest angular resolution, as a continuation of our results on rotation measurement gradients across the jet. This is the main goal of one of the RadioAstron Key Science Projects, for which observations will start in September 2013.
- Jet wobbling. We have past and ongoing observations to map and monitor the innermost total flux and polarimetric jet structure in NRAO150 and other sources with the GMVA and complementary 7mm VLBA observations, aimed at determining a jet swing model that better fits the observational evidence (precession, orbital motion, or jet perturbation).

*C11: Imaging the base of the jet in M 87*
K. Hada[26], G. Giovannini[26,14], M. Giroletti[26], M. Orienti[26]

Our research group is strongly interested in observing the nearby radio galaxy M 87 with a global mmVLBI array including ALMA baselines, especially at 3mm or 7mm. Because of its proximity, M 87 provides the best opportunity to probe the ultimate formation mechanisms of black-hole-powered jets. Our recent VLBA study has demonstrated that the jet formation region within a few tens of Schwarzschild radii is possible to image for M 87 at mm wavelengths. The crucial next step is to observationally constrain the detailed collimation and acceleration structure in this region, and then perform direct comparisons with theories. To realize this goal, we strongly stress the importance of global-VLBI array at 3mm/7mm including the phased-ALMA, because this will dramatically improve the imaging ability for the base of the M 87 jet in terms of both sensitivity and resolution.

*C12: The structure of AGN on Event Horizon Scales and the Origin of Jets*
J.A. Zensus[44,51], T. Krichbaum[44], A. Lobanov[44], A. Roy[44], A. Eckart[51,44], M. Bremer[29], I. Martí-Vidal[48]

The use of the phased ALMA will bring mmVLBI to new frontiers and offer the sensitivity that is needed to image in detail the innermost regions in AGN jets with unprecedented angular and spatial resolution. The MPIfR-IRAM-OSO team has a well-established legacy in mmVLBI related activities and work, both scientific and technical. The group, consisting of several senior scientists, postdocs and numerous PhD students, therefore is well prepared and has a strong interest in the continuation of this research, and in pushing the observing technologies further. Specific research interests are:
- Imaging the emission around nearby super massive black holes and on event horizon scales. This includes a detailed study of Sgr A* and M 87 using mmVLBI with ALMA at all available frequencies and in polarization. It also invokes time resolved monitoring observations to search for structural variability and for tests of Einstein's General Relativity and the measurement of the spin of the black holes.
- In collaboration with Cologne University (A. Eckart) we intend to perform triggered VLBI imaging of Sgr A*, using NIR variability as trigger for immediate VLBI imaging at the highest possible frequencies.
- Study of the jet launching and acceleration zone in nearby AGN of which about 20 can be imaged with a spatial resolution of better than 500 Schwarzschild radii aimed at the better understanding of how jets are made.
- Systematic measurement of the brightness temperatures in complete samples of compact radio sources. The knowledge on the frequency dependence and on the radial dependence of TB will allow discriminating between different jet launching models and on the internal composition of jet near origin.
- Imaging AGN at times when they are active or flaring which will allow relating jet kinematics to the broad-band variability, helping to understand high energy emission mechanisms (X-ray, gamma-rays, TeV) and in AGN jets.
- Trace bent trajectories of superluminal jet components in AGN jets back to origin and answer the question of their physical cause (e.g. binary black hole precession vs. instability interpretation).
- Image faint radio galaxies at high redshifts to study effects of cosmological evolution.

*C13: Parsec-scale environment of AGN*
M. Orienti[26], M. Giroletti[26], G. Giovannini[26,14], F. D'Ammando[26], R. Lico[26], D. Dallacasa[14], T. Venturi[26], M. Bondi[26], I. Prandoni[26]

The excellent sensitivity achieved by mmVLBI with a phased ALMA will enable a deep look into the innermost part of the jet that can reveal the acceleration and collimation region very close to the black hole (<100 Schwarzschild radii) where magneto-hydrodynamic instabilities form. Simulations suggest changes in the jet collimation angle with the distance from the central engine and mmVLBI can provide a direct observational probe of this region. The high frequency observations with the sub-pc resolution will be a major breakthrough in the study of the region responsible of the high-energy emission. The addition of the polarization information will allow us to constrain the structure of the magnetic field as well as the shock scenario, thanks to the low-effective Faraday rotation in these objects. The detection of compact, high-brightness temperature regions and their location along the jet may solve the debate on where the high-energy emission is produced and the main radiative processes involved.

*C14: Core-shifts and opacity effects in quasar radio cores*
E. Ros[44,4,12], J. Guirado[12], M. Kadler[30], M. Perucho[12], C. Fromm[44]

The study of the opacity at the base of the jet in AGN, and this is a topic where the sensitivity and resolution provided by ALMA will be of special importance. Addressing the position of the jet base with VLBI has been performed either by astrometric methods or by aligning the images to common, optically thin features present at different wavelengths. mmVLBI can explore the jets at the innermost region and probe the size of this area and the potential location of the black hole with respect to the core and putative recollimation shocks close to the core, or changes of jet position angle with frequency revealing a helical nature, or even time-changing orientation of the
mm-wavelength jet to be related to precession of the relativistic jet.

*C15: The relation between gamma-ray and mm-wave emitting regions in Fermi sources*
E. Ros[44,4,12], M. Kadler[30], J.A. Zensus[44,51]

To complement the TANAMI collaboration, where we are monitoring 80 sources at 8.4 and 22 GHz. Several TANAMI sources are key objects at the southern latitudes and can be addressed with a large southern antenna. Furthermore, intermediate latitude, strong sources such as 1921-293, 0607-157, 1730-130, 1334-127, 1253-055, 0420-014, and included in other samples, are also of special interest. Not to forget the relationship between the radio properties with the gamma-ray emission, which has been studied extensively in recent times after Fermi's launch.

*C16: Magnetic field structure and the acceleration region in radio jets*
T. Savolainen[44] and MPIfR VLBI-group

Millimetre-VLBI is the only method to directly image the AGN jets down to their formation region and to study the acceleration and collimation zone in nearby AGN in the optically thin regime. Such observations can provide clues about:
- Whether the behaviour of the flow follows the qualitative predictions of the current state-of-the-art MHD jet acceleration simulations, in which tremendous progress has been made over the past decade.
- Where the transition from the Poynting-flux-dominated to kinetic-flux-dominated flow takes place.

- Whether the transition is linked to the energy dissipation in the jet and to the high-energy emission production in blazars.
- Whether the transition takes place in similar way in all classes of jets.

Polarimetric VLBI observations at 0.8 and 1.3 mm can directly probe the magnetic field structure of the acceleration and collimation one. At the end of the acceleration zone a rapid pinching may generate a standing recollimation shock, a potential site for producing the variable non-thermal continuum emission of blazars, the "VLBI" core, and the source of high-energy gamma-ray emission. Multi-frequency and polarimetric mmVLBI can resolve this structure and determine if the "core" indeed is a physical feature, instead of just being the location where the jet becomes optically thin. This is because a recollimation shock is expected to have a frequency-independent position and a distinctive polarization structure.

*C17: Detection and evolution of supermassive binary black holes*
S. Britzen[44], A. Eckart[51,44], L. Rezzolla[31,45], O. Kurtanidze[1]

The cosmic growth and evolution of supermassive black holes is far from being understood. Galaxy collisions and the subsequent merger of the black holes at their centers may play an important role in triggering the active phase of AGNs. Yet, identifying supermassive binary black hole (SMBBH) systems on pc-scales has proven difficult. The most promising results were obtained by deducing the existence of these systems on the basis of the theoretical modelling of optical light-curves or VLBI jet monitoring data. A beam-formed ALMA as part of a global mmVLBI may allow to directly detect the close binaries, study their evolution through cosmic times and their role with regard to AGN activity and star formation (e.g., in ULIRGs). ¬†This area of research will gain even more attention with the future possibility of gravitational-wave detection by Pulsar Timing Arrays and the upgraded gravitational-wave interferometers.

*C18: Merging Black Holes -- from encounter to coalescence*
H. Messias[10,63], S. Cales[63], J.M. Afonso[10]

A strong connection between actively accreting nuclear black holes and their host (dynamical/luminous) mass has been observed to hold in a range of up to ~3 decades in black hole mass. In a hierarchical galaxy-evolution scenario one thus expects the occurrence of nuclear black-hole pairs as a result of host galaxy mergers. However, the evolution of a black-hole pair system from initial merger distances (~100's kpc) to distances where gravitational radiation is expected to dominate (<~0.01 pc) is anything but trivial. Three-body interaction with the stellar environment and dynamical friction with the host gaseous content are expected to be mechanisms which induce black-hole pair hardening, i.e. drive the system to become more gravitationally coupled/bounded. Our understanding of massive black-hole pairs ($>\sim 10^7$ $M_{sun}$), is still poor in explaining how they can merge within a Hubble time. If indeed they do not merge, one would expect to still be able to observe back-hole pairs in the most massive galaxy hosts. With an angular resolution as low as tens of micro-arcsec, a mmVLBI facility can probe pair-distance scales down to the gravitation dominated regime (~0.01pc) up to redshifts of ~0.034.

In spite of considerable effort over the past decade, the task of finding binary back-hole systems has proven to be demanding. On the one hand, direct evidence of black hole pairing (e.g. via resolved X-ray and radio imaging) is limited to a phase where the black-holes are still gravitationally unbounded, while closer binaries are difficult to confirm given the ambiguity of the adopted techniques (i.e. a for double peaked high-ionization emission lines). A mmVLBI facility will allow for a detailed view of gas dynamics in the closest regions to the black-hole pair and, in some cases, jet production. Such observations will provide key constraints not only for initial conditions, but also for more evolved stages of black-hole pair evolution (e.g. prograde or retrograde systems). The synergy with micro-arcsec optical-to-near-IR imaging (possible with e.g. JWST) and/or IFU studies (a growing capability in major observational facilities with adaptive-optics systems) of the host stellar content will complement the study of the possible mechanisms driving black-hole pair hardening.

*C19: Missing flux in SiO masers around AGB stars*
F. Colomer[47], V. Bujarrabal[47], J. Alcolea[47], J-F. Desmurs[47], R. Soria-Ruiz[47], A. Richards[38], M. Lindqvist[48], H. Olofsson[48], W. Vlemmings[48], E. Humphreys[15], A. Baudry[7]

VLBI and multi-line observations of SiO masers have provided extremely valuable information on the inner circumstellar shells around AGB stars, and on the pumping mechanisms responsible for the emission in these regions. VLBI observations allow us to map the brightest SiO spots in late-type stars; these spots correspond to the highest maser amplification paths. However, it is clear that most of the SiO maser flux that is seen in single-dish spectra is not detected in VLBI long baselines. This missing flux could be due to an unseen distribution of weak maser clumps and/or to the existence of an extended emission across the circumstellar shell. Even if that would be true (because of spatial over-resolution), a larger sensitivity will result in the detection of much more emission. Checking the complete distribution of the maser emission is basic to develop pumping models, including weaker emission features, to better probe these inner regions of the circumstellar envelopes. Multi-line VLBI observations at 7 and 3 mm are proposed with participation of two or three phased ALMA sub-arrays.

*C20: Galactic dynamics: parallaxes and proper motions of SiO maser stars*
F. Colomer[47], V. Bujarrabal[47], J. Alcolea[47], J-F. Desmurs[47], R. Soria-Ruiz[47], M. Lindqvist[48], H. Olofsson[48], W. Vlemmings[48], H. van Langevelde[39,2], A. Baudry[7], L. Chemin[7], P. Charlot[7], G. Bourda[7]

In the radio domain there is considerable ongoing effort to measure parallaxes and proper motions of stars exhibiting SiO emission (e.g. VERA interferometer observations in Japan). Phased ALMA will greatly enhance such studies and, together with other arrays or telescopes (ouch as EVLA, VERA, ATCA and KVN), will provide unprecedented sensitivity for high precision measurements of parallaxes and proper motions of SiO stars in the Galactic center and arms by means of the cluster-cluster techniques (sub-arraying each instrument). There are hundreds of SiO stars available in the southern hemisphere to complete this task, investigate the complex dynamics of stars around the Galactic center, and bring new constraints on the Galactic rotation curve. The current successful efforts to measure parallaxes and proper motions in SiO-emitting stars guarantees the feasibility of such studies in obscured regions of the Galaxy where only the cool CO gas was mapped. Results will fruitfully be compared with dynamics studies performed in the optical domain with the Gaia space project.

*C21: Multiline-line maser observations at very high frequencies*
F. Colomer[47], V. Bujarrabal[47], J. Alcolea[47], J-F. Desmurs[47], R. Soria-Ruiz[47], M. Lindqvist[48], H. Olofsson[48], W. Vlemmings[48], J. Cernicharo[8], M. Gray[38], A. Richards[38], E. Humphreys[15], A. Baudry[7], F. Herpin[7]

Several ALMA bands (Bands 4, 5 and 6) will allow us to observe a large number of SiO maser transitions in late-type stars, still unexplored with VLBI baselines. These SiO transitions are not observable with other VLBI instruments, such as the VLBA, EVLA, VERA, or EVN. The mechanisms and conditions able to explain such high-excitation lines are very poorly understood, although complex processes, such as interaction between lines of different species, are suspected to be involved. The existing observational information on these lines will be renewed with new high-resolution mmVLBI data, supposing that in the future there will be VLBI instruments other than ALMA able to observe these lines (e.g. IRAM interferometer operated in cluster-cluster mode to ease phase calibration). Comparison of the brightness distributions of the high-frequency lines will give a more complete picture of the dynamics and excitation conditions in evolved stars.

*C22: Prospective studies of HCN masers in Carbon-rich stars*
F. Colomer[47], V. Bujarrabal[47], J. Alcolea[47], J-F. Desmurs[47], R. Soria-Ruiz[47], H. Olofsson[48], W. Vlemmings[48], E. Humphreys[15], A. Richards[38], A. Baudry[7], F. Herpin[7], K. Menten[44]

HCN submillimetre and millimetre maser emission has been detected from carbon-rich circumstellar envelopes (with [C]/[O] photospheric abundance ratios > 1) where masers from other species such as SiO, $H_2O$ and OH are not present. Brightness temperatures > 10000 K have been measured using connected interferometry, but the emission distribution remains unresolved. In the past, VLBA observations of CIT 6 were unsuccessful mostly because of the relative weakness of the emission in comparison with other circumstellar masers; no further VLBI studies have been performed toward other carbon stars.

The enhanced sensitivity of a mmVLBI array including the ALMA phased array would allow, for the first time, to study the distribution of HCN masers and would provide the most accurate view of the dynamics of the innermost regions of these circumstellar envelopes where molecules are being formed and the mass-loss is being driven (similarly to what occurs with the SiO masers in O-rich stars)

*C23: Sub-millimetre H20 masers in late-type stars*
A. Richards[38], M. Gray[38], E. Humphreys[15], A. Baudry[7], F. Herpin[7], F. Colomer[47], V. Bujarrabal[47], J. Alcolea[47], J-F. Desmurs[47], R. Soria-Ruiz[47], H. Olofsson[48], W. Vlemmings[48], J. Brand[26], S. Etoka[58], K. Assaf[38], K. Menten[44]

Several projects for mmVLBI of water masers with a subset of ALMA phased antennas (or with all antennas) seem possible in the near future in combination with ALMA 'connected' interferometry. We have identified 3 main subjects of research. Several evolved stars could be adequate targets. Over a dozen mm/sub-mm lines have now been detected (see e.g. Neufeld et al. 2013, ApJ 769, 48N). We consider the 321 and 325 GHz transitions likely to have the best signal-to-noise on VLBI scales. Other transitions may also be accessible but hardest to observe (e.g. the strong 183 GHz transition, except when it is red-shifted away from the atmospheric line, as is the case for megamasers). Questions to be addressed include:
- How does mass lost from the stellar surface reach escape velocity? At >~5 stellar radii, the winds are driven by radiation pressure on dust and cm-wave masers show steady radial acceleration. Closer to the star, SiO masers show both infall and outflow. mm and sub-mm water masers are the only species likely to probe both regions. Line widths and statistics from single dish observations (e.g. Cernicharo et al. 1990, Yates et al. 1995, 1996) and models (e.g. Menten & Melnick 1991, Humphreys et al. 2001) suggest that the 325 GHz transition would peak outside the dust formation radius and the 321 GHz transition inside, but this is yet to be confirmed observationally. VLBI imaging will show which lines are segregated and which are associated on scales of 0.001 to 0.1 AU. This is the only way to constrain the molecular composition and physical conditions on scales small enough to investigate whether clumps ejected from the star have distinct physics and chemistry, as well as tracing changes in kinematics across the dust formation zone.
- Maser beaming - a shock diagnostic. Masing clouds around AGB and RGB stars have typical radii ~ 10 mas. Individual 22-GHz maser spot sizes vary from microarcsecs to the size of the masing cloud. Very small apparent sizes are expected due to maser beaming from approximately spherical clouds, but shocked slabs can produce extended components, which are indeed seen around stars likely to have the deepest pulsations. Comparing VLBI and mas-scale observations of sub-mm masers will reveal whether shocks have a stronger impact nearer the star (thus, individual 321-GHz maser spots would be more extended than 325-GHz).

- Turbulence. Proper motions of maser spots will demonstrate a combination of infall, outflow, possibly rotation (although this has never been confirmed around solitary stars) and turbulence; the last of these can only be measured directly on VLBI scales. Fractal analysis of maser spot clustering measures turbulence scales, related to compressibility and to any physically distinct structures in the wind. This will distinguish maser patterns as accidents of beaming due to turbulence, from physically discrete clouds with different properties.

*C24: Gas dynamics and physical conditions in massive protostellar outflows*
C. Goddi[39], H. van Langevelde[39,2], L. Moscadelli[27], G. Surcis[39], E. Humphreys[15], W. Vlemmings[48], J. Brand[26], F. Herpin[7], A. Baudry[7]

Multi-epoch VLBI observations enable to accurately measure the 3D dynamics of circumstellar gas in star-forming regions (SFRs). Besides the ubiquitous 22 GHz line, strong $H_2O$ lines have been detected at 183, 321, and 325 GHz in SFRs using single-dish telescopes. Their spectra show spectacularly intense peak fluxes (up to 10000 Jy) and large velocity ranges (up to 300 km/s). Observations of maser line ratios may be used in principle to constrain physical conditions, but mmVLBI observations of maser lines are needed to map the gas temperature and density with high accuracy and resolution in the vicinity of YSOs. So far, only two SFRs, Orion-KL and Cep-A, have been imaged in the sub- mm $H_2O$ lines with the SMA. The inclusion of ALMA in a mmVLBI network is mandatory to enable studies of sub-mm $H_2O$ maser lines in a larger sample of SFRs and a direct comparison with the cm lines observed at 0.001 arcsecond resolution. The main goals of VLBI observations of mm water maser lines are:
- To map temperature and density distributions of the circumstellar gas at very high spatial resolution by using radiative transfer models constrained by the sub-mm and cm line data;
- To map gas dynamics in protostellar outflows at very high spatial resolution (at AU- scales eventually showing fine details of the protostellar disk and/or of the base of the associated outflow).

*C25: Magnetic fields in the circumstellar gas at small radii from massive YSO's*
C. Goddi[39], H. van Langevelde[39,2], G. Surcis[39], W. Vlemmings[48], J. Torrelles[36], B. Hutawarakorn Kramer[43]

VLBI observations of maser polarization conducted in the last few years have enabled to reveal unique information on the magnetic field strength and structure in high-mass SFRs. $H_2O$ measurements give magnetic field strengths of order 100 mG at densities of about 10(9) cm-3 and physical scales of less than 1000 AU, reveal magnetic structures in outflows, and even provide evidences of different protostellar evolutionary stages. Once ALMA full polarization capabilities are offered, it will be possible to conduct detailed studies of polarized maser emission towards SFRs also using mm maser lines. For example, one advantage of observing maser polarization at different frequencies and/or resolution towards an extended source is the possibility to constrain the magnetic field properties, like field strength and/or direction, in different portions of the circumstellar gas within the observed source. The VLBI capabilities of ALMA will overcome the blending of differently polarized features, especially in the very complex protostellar environment.

*C26: Distances and 3D velocities of YSOs*
C. Goddi[39], H. van Langevelde[39,2], L. Moscadelli[27]

Parallaxes of masers are an excellent method to measure distances and 3D motions of massive YSOs. At cm-wavelengths, there is a considerable effort to measure parallaxes and proper motions of YSOs exhibiting water masers (22 GHz) and methanol masers (6.7 and 12.2 GHz) in our Galaxy. Phased ALMA would provide unprecedented sensitivity for high-precision measurements of parallaxes and proper motions of SFRs associated with strong maser emission in the southern hemisphere, thus bringing new constraints on the Galactic rotation curve. High-precision absolute astrometry could be achieved using two ALMA sub-arrays (each one phased), targeting simultaneously the water maser source and a nearby quasar.

*C27: Unveiling the powering source of the nearest massive SFR: Orion BN/KL*
C. Goddi[39], E. Humphreys[15], F. Niederhofer[15]

SiO masers have provided a unique perspective in the making of a massive star, Orion Source I, the closest known example of massive YSO. Extensive VLBI imaging of SiO masers at 7mm (J=1-0) and 3mm (J=2-1) revealed for the first time the launch and collimation region of an outflow from a rotating compact disk on scales comparable with the Solar System. The fact that all the flux is recovered at the VLBI resolution at 43 GHz (0.5 mas), suggests that mmVLBI observations with phased ALMA will likely be successful. This would enable more complete mapping of the circumstellar material at radii 1-100 AU from the central YSO, providing unprecedented details on the 3D gas dynamics in HMSF. Using ALMA-Cycle 0 Band 6 observations, SiO transitions from the J=5-4 rotational levels have been imaged finding first evidence of a high-frequency SiO maser from the vibrationally- excited state (v=1). New high-resolution data of these high-excitation lines will enable us to better constrain the relatively poorly understood excitation mechanisms of these SiO masers lines.

*C28: A celestial reference frame at high frequency to study AGN core-shifts*
P. Charlot[7], G. Bourda[7]

Our aim is to use the mmVLBI array including ALMA to build a celestial reference frame at high frequency (86 GHz at least). The goal is to study the frequency dependence of the quasar positions that are used to establish such a frame. The current celestial frame is the International Celestial Reference Frame (ICRF), which is based on observations conducted at the standard "geodetic" frequency of 8 GHz with simultaneous observations at 2 GHz to correct for ionospheric effects). The ICRF includes 3400 sources with a floor in coordinate accuracy of 40 microarcseconds. Extensions are also being made at 22, 32 and 43 GHz. By 2015-2020, the Gaia space astrometric mission (to be launched in the fall of 2013) will produce a QSO-based optical celestial frame with similar accuracy while including many more objects. Establishing a celestial frame at 86 GHz (or above) with the ALMA-VLBI facility would provide quasar positions at a frequency intermediate between that of the current ICRF and the Gaia optical frequency. Such a frame would thus be of high interest to determine the relative location of the emitting regions at cm, mm and optical wavelengths (the so-called core-shifts) and thereby help to understand AGN physics.

*C29: Testing the unified AGN model in local AGN*
E. Humphreys[15], I. di Gregorio[15], J. Brand[26], A. Tarchi[28], P. Castangia[28]

Extragalactic masers at low frequency (< 100 GHz) are found in a number of environments including Active Galactic Nuclei (AGN), and starburst galaxies. 22 GHz water masers are found in AGN nuclear disks (r < 1 pc) and outflows (r < 10 pc). More than 150 water maser galaxies have now been detected thanks to efforts such as the Megamaser Cosmology Project. A gravitationally lensed water maser has also been detected at a redshift of z=2.64. Masers currently provide the only way to map the structure of circumnuclear accretion disks within a parsec of supermassive black holes (SMBHs). In local AGN (D < 30 pc), mmVLBI observations of masers can be used for testing the unified model for AGN (e.g. is there a need for a torus) and AGN central engine physics. Via a combination of spectral monitoring of the masers to determine centripetal accelerations, mmVLBI mapping of the masers, and modelling of the nuclear disk, geometric distances to host disk maser galaxies can be obtained. In addition to this, high-accuracy mass estimation for SMBHs can be made to place the maser galaxies on the M-sigma diagram. By determining disk warps and sub- structure (e.g. spiral structure in the NGC 4258 disk), AGN central engine physics and the AGN unified model can be tested.

*C30: Maser cosmology and determination of the Hubble constant*
E. Humphreys[15], I. di Gregorio[15], J. Brand[26], A. Tarchi[28], P. Castangia[28]

For distant AGN (> 50 Mpc), maser cosmology can be used for the determination of a high accuracy Hubble Constant and thereby better constraint of the Dark Energy equation of state parameter w. Outstanding issues remain at 22 GHz. The M-sigma diagram is poorly sampled by 22 GHz water maser galaxies which occupy a relatively narrow range of SMBH mass, AGN disks are poorly sampled by 22 GHz masers i.e. the back of the disk has never been detected in low velocity systemic emission, perhaps due to obscuration by ionized material above the disk or the jet core at 22 GHz. Since the determination of Ho relies on imaging, we need to find stronger/unobscured masers to perform cosmology using more distant maser galaxies. Hence the search for extragalactic mm/sub-mm masers. At high redshifts, arguably more promising maser lines can become shifted into regions of good atmospheric transmission. This applies to 183, 321 GHz and 325 GHz. These masers could all be useful for the distant AGN cases, and we may also be able to probe very highly red-shifted gravitationally lensed sources. The geometric distances to individual galaxies are relatively high-accuracy with the potential of delivering a high-accuracy Hubble Constant from a collection of maser galaxies selected to sample the Hubble Flow.

*C31: Magnetic field mapping of the near stellar environment*
W. Vlemmings[48], E. Humphreys[15], M. Gray[38], A. Richards[38], F. Herpin[7], A. Baudry[7]

Highly polarized (>20%) high-frequency SiO masers have been shown to be good probes of the magnetic field morphology close to the stellar atmospheres, within a few stellar radii. With mmVLBI including a phased ALMA it will then be possible to map in detail the magnetic field on the smallest scales by using a variety of different maser species and transitions. In subsequent monitoring observations, it could become possible to trace magnetic ejections from evolved stars and directly determine their magnetic activity. It should also be possible through a number of tracers, SiO being the tracer on the smallest scales, to link stellar fields to larger scale magnetic fields in the interstellar medium.

Current magnetic field studies in late-type stars have by necessity been focused on the oxygen rich stars, but mmVLBI gives access to the 89 GHz HCN maser, which could allow similar studies to be performed in carbon-rich envelopes.

*C32: Hydrogen recombination lines, masers, and non-thermal emission from young-stellar objects*
R. Galvan-Madrid[15]

Research interest in high-resolution observations and modelling of hydrogen recombination lines (RLs), masers, and non-thermal emission from young-stellar objects. Some RL objects appear to be photoionized disks around young stars, which can mase at mm wavelengths (e.g. Weintroub et al. 2008, ApJ, 677, 1140). This emission is on scales < 10 mas and has not been angularly resolved. Also, if bright enough, the periodic synchrotron emission of magnetic reconnection events in binary young stars could be mapped, since it is expected to arise at scales <0.1mas (e.g. Salter et al. 2010, A&A, 521, 32).

*C33: High redshift galaxies and Masers in Nearby Galaxies and AGNs*
E. Liuzzo[26], V. Casasola[26], M. Massardi[26], A. Mignano[26], I. Prandoni[26], R. Paladino[26], J. Brand[26]

- High redshift galaxies. Including ALMA in mmVLBI observations will provide the sensitivity and angular resolution required to study scales of the order of pc in high redshift galaxies ($z>2$). mmVLBI+ALMA continuum and spectroscopic observations will allow:
    - To investigate the mechanisms of inflow operating in the regions surrounding the super massive BH in distant AGN
    - To separate star formation from AGN emission in high-redshift galactic nuclei
    - To provide BH mass estimates from the molecular gas dynamics on pc scales as a function of redshift;
    - To investigate merging effects and the structure of the dust tori as a function of redshift
    - To map the dust-unobscured gas structure in high redshift 'normal star-forming galaxies, possibly exploiting weak lensing.
- Polarization. High frequency full-Stokes investigation of AGN will be capable to retrieve a Faraday-rotation independent description of the structure of the jets, tracing the magnetic fields down to their basis. The high frequency small beams reduce the effects of beam depolarization. As polarized emission is typically only a fraction of the emitted signal, it is crucial to improve the mmVLBI array sensitivity by including ALMA in the array.
- Masers. Observations of sub-mm maser emission in regions of star formation, to trace the (kinematics of) accretion disks and/or the base of the jet originating at the protostar which drives the larger-scale outflow, and in circumstellar envelopes around evolved stars to trace the dust formation zones and the dynamics of the mass loss process.

*C34: mmVLBI of water masers*
A. Tarchi[28], P. Castangia[28]

Our research activity is focused on the search, study, and analysis of sources of the main water maser line at 22 GHz hosted in active (starburst and AGN) galaxies. The possibility to perform VLBI high-resolution measurements of mm/sub-mm water maser transitions, as found in NGC3079 and Arp220 and, more recently, in Circinus Galaxy, will provide a fundamental complementary tool to obtain physical and kinematic parameters of star formation and active nuclear regions in nearby (although, not exclusively) galaxies. This becomes particularly relevant given our involvement in the construction and operations of the upcoming Sardinia Radio Telescope, that it is expected to work up to mm wavelengths and be part of the Global mmVLBI network.

*C35: Disks of nearby stars*
R. Greimel[22], M. Leitzinger[22], M. Temmer[22], P. Odert[22], R. Brajsa[21], A. Hanslmeier[22]

The high spatial resolution of a proposed future global mmVLBI facility enables resolved measurements of the disks of nearby stars. In particular it would facilitate the investigations of active stellar regions, their surface distribution, temporal evolution, and connected transient phenomena. This would allow the measurement of stellar rotations, activity cycles, spot locations (pole or equator), and their dependence on stellar spectral type and age. Interesting new discoveries are expected from the T-Tauri to the subgiant phase of stellar evolution with new mm observing capabilities.

*C36: Jets in Protostars*
P. Klaassen[2]

- Based on recent models, an accreting 'high-mass' star will puff up when it gets above a few solar masses because of entropy within the star. This happens for high, but realistic, accretion rates and once the star gets to about 10 $M_{sun}$, its radius will be about 100 $R_{sun}$ (0.47 AU or 0.4 mas at 1 kpc distance).
- Protostars, similar to AGN, will also produce jets. They will of course be orders of magnitude less powerful, but the protostars are also orders of magnitude closer. For a ~ $M_{sun}$ star, the jet launch radius is 0.1 AU (0.1 mas at 1 kpc). These size scales are in the regime of mmVLBI, although the sensitivity may not be sufficient.

*C37: Extragalactic absorption lines in quasar tori*
V. Impellizzeri[69], T. Wiklind[66], A. Roy[44]

At sub-millimetre wavelengths, absorption systems towards distant quasars can tell us about the star formation history as well as abundances and initial conditions for the ongoing star formation. The evolution of chemical abundances can be studied through the detection of a large variety of molecules detected in the diffuse gas. Furthermore, absorption studies can be used measure the CMB temperature as a function of redshift, estimate the Hubble constant independently through time delay between gravitational lensing, and probe the possible variation of fundamental constants.

All these studies require high angular resolution to resolve the absorbing gas, than is unavailable by current sub-mm interferometers. Without detailed knowledge of the small-scale structure of the molecular gas, however, we can only derive lower limits to fundamental parameters such as the column density. A high sensitivity and resolution mmVLBI array can constrain the smallest scales of the molecular gas by looking for variations in the column density as a function of time, as seen in the gravitationally lensed QSO PKS1830- 211 as well as PKS14130+1413. Towards the core of active galaxies, we will be able to peer into the so far elusive molecular
torus region, thereby testing the physical conditions around supermassive black holes, and complementing all the work done on and about the event horizon, jet launching and variability in the core.

*C38: Molecular absorption against radio quasars*
S. Muller[48], F. Combes[41]

We are interested in the phasing of ALMA in the global mmVLBI network for our work on high-redshift molecular absorption against radio quasars. Being distant independent and provided the background continuum source is strong enough, there is no sensitivity problem, to detect molecules, even in very small quantities down to a solar mass. The gain in sensitivity with ALMA phasing, combined with the high resolution of VLBI, would provide us with a better understanding of the molecular absorbers. In turn, molecules can be used as key diagnostics tools of the gas properties and as cosmological probes. In particular, we would like to perform VLBI studies of molecular absorbers to:
- Explore the chemical properties of clouds at intermediate redshift and characterize the type of absorbing clouds
- Explore the physical properties of clouds in particular the small-scales power distribution. Here the statistics can be potentially boosted by the variability of background sources (implying VLBI monitoring).
- Constrain the cosmological variations of the fundamental constants of physics, such as the fine structure constant and the proton-to-electron mass ratio. When comparing lines from different species, the high resolution of mmVLBI is key to ensure that the species are co-spatial, and minimize confusion by kinematical/chemical shift.

- Explore the evolution of the CMB temperature as a function of redshift. Here again the VLBI resolution is key to ensure that the molecules/lines used as thermometers are co-spatial.
- Refine the lens model when the background quasar is lensed by the intervening galaxy. mmVLBI offers the best angular resolution. Resolving substructures in the lensed images (core/jet) give strong constraints on the lens model.

*C39: High-peaked BL Lac objects*
M. Kadler[30], K. Mannheim[30], R. Schulz[30], A-K. Baczko[30], C. Mueller[30], D. Eisenacher[30], A. Kreikenbohm[30], J. Truestedt[30], J. Wilms[56], E. Ros[44,4,12], T. Krichbaum[44], M. Boeck[44], J. Hodgson[44], L. Fuhrmann[44], J.A. Zensus[44]

- We are experienced users of VLBI facilities (GMVA, EVN, VLBA, LBA+). In the mm-regime, we have conducted several successful GMVA and VLBA projects. NGC1052, the subject of a PhD project, is one of the closest (~20Mpc) and faintest (~400mJy) AGN ever successfully imaged with VLBI at 3mm wavelength. Another PhD project concentrates on the possibly closest known blazar, IC310, which is currently beyond the sensitivity limits for mmVLBI studies and would a prime target for future mm- VLBI arrays including a phased ALMA.
- With reference to Ros' contribution (C14): we are coordinating the TANAMI project from Würzburg (in collaboration with Roopesh Ojha at GSFC). This is the largest ongoing VLBI monitoring program in the southern hemisphere. Only the northern- most TANAMI sources can be observed with the current GMVA. With a phased ALMA and other future southern-hemisphere telescopes, it would be possible to go to more southern declinations and the enhanced sensitivity would open the possibility to observe fainter objects such as the high-peaked BL Lac objects in our sample.  Most projects in our team are set up as multi-wavelength programs, involving coordinated optical/UV, X-ray, gamma-ray and VHE observations.

*C40: Pulsars studies with a phased ALMA: pulsars around Sgr A\*, emission physics, and neutrons star populations*
M. Kramer[44,38], R. Eatough[44], J. Verbiest[18], N. Wex[44], B. Stappers[38], H. Falcke[23,44], L. Rezzolla[31,45], J. Lazio[68], S. Ransom[69], J. Cordes[62]

We would like to express our interest in using a phased-up ALMA for a number of science topics. These include:
- Searching and monitoring pulsars orbiting SGR A\*: Searching and monitoring pulsars orbiting SGR A\* will enable to us to measure the properties of SGR A\* to unprecedented precision and to test GR's descriptions of BH.
- Pulsar emission physics: One of the greatest challenges in astrophysics is in understanding the mechanism that creates coherent pulsar emission. Going to very high radio frequencies has the potential to unveil the coherence length intrinsic to the process. Understanding the emission process is ultimately also important to infer the intrinsic limits of using pulsars as clocks.
- Neutron star population: In recent years, magnetars have been discovered at radio frequencies. It turns out that in contrast to radio pulsars, this special class of neutron stars shows a very flat flux density spectrum. Phased-up ALMA observations would allow us to probe this small but important part of the Galactic neutron star populations and to study the difference to radio pulsars.

*C41: Beyond the Standard Model: constraining the hidden sector of weakly interacting and ultra-light particles*

A. Lobanov[44], H. Zechlin[60], D. Horns[58]

Most extensions beyond the Standard Model (SM) predict the existence of a "hidden" sector of weakly interacting and ultra-light particles. In particular, WISP axions and hidden photons, which are considered among the best-motivated dark matter particle candidates, can be best searched with laboratory and astrophysical measurements in the radio through gamma-ray regimes. High-sensitivity ALMA measurements of broadband spectra of compact radio sources such as outer planets, supernova remnants and active galactic nuclei will be instrumental for the hidden photon searches at extremely low energy/mass scales. ALMA observations of AGN, SNR, and outer planets will offer an excellent tool for placing bounds on the mixing angle. The effective sensitivity of ALMA measurements will depend on the line-of-sight density profile of the interstellar and intergalactic medium, producing a resonant enhancement of the sensitivity at frequencies close to the plasma frequency and causing a strong damping of the oscillation signal at lower frequencies. ALMA measurements would provide a unique probe for large areas of the parameter space even in a worst-case scenario of the plasma distribution.

*C42: VISTA Variables of the Via Lactea Survey (VVV): embedded variables and the search for hidden Supernovae*
D. Minniti[33,61], P. Lucas[59], M. Rejkuba[15], N. Masetti[24], V. Ivanov[66], N. Cross[53], E. Kerins[54], M. Thompson[59]

The VISTA Variables of the Via Lactea Survey (VVV) is an ESO public near-IR variability survey that is sampling the Galactic bulge and an adjacent section of the Southern Galactic plane. The bulge hosts crucial stellar populations encompassing the Galactic center, whereas the inner disk features regions of intense star formation around spiral arms. The confirmation and follow-up of sunset of variable objects with ALMA is an essential component of the project. As a first relevant example, a search of the VVV 1st Data Release led to the discovery of VVV-WIT-001, a very high amplitude variable with unusual colors from amongst 50 million sources. VVV-WIT-001 is very red: H-Ks=5.2 mag, and invisible in the Z, Y and J passbands. The very rapid decline in flux by >6.5 magnitudes plus the very red colors are unprecedented, and inconsistent with known examples of variable stars. Therefore, this object would have to be the prototype of a new sub-class of PMS variable, or an even more exciting possibility, it could be a highly obscured supernova in the Milky Way, detected at some time after maximum light. This source is now being followed up with ALMA observations in order to confirm or discard a new SN in the Milky Way. This is surely the first of many embedded variable targets for which ALMA observations would be crucial to reveal their unknown origins and physical mechanisms.

*C43: Jets from X-ray binaries*
A. Rushton[55], D. Altamirano[3], M. Coriat[65], R. Fender[55], E. Kording[23], S. Markoff[3], S. Migliari[11], M. Felix[5], J. Miller-Jones[67], Z. Paragi[39], V. Tudose[32], G. Sivakoff[70], R. Soria[67], R. Spencer[38]

X-ray binaries (XRBs) are an important class of astronomical object for studying the physics of black hole accretion. Due to their similar morphology to quasars, albeit on much smaller size scales, jet forming XRBs are also known as `microquasars'. Given that XRBs behaviour can vary in timescales of weeks to months, they can be used as proxies for understanding AGN behaviour that occurs on times scales much greater than a human life time. Moreover, we do not understand how jets are launched and collimated. For example, we know that XRBs can produce `steady' jets that can be replaced by discrete ejection of plasma, but we don't know the exact details of what happens. The inclusion of ALMA into a Global mmVLBI network will enable the study of many known XRBs at <1 mas and scales of ~10^(9-10) meters or the inner 0.015-0.15 AU of the binary. Such an instrument would give light into the following XRB topics:

- Very high-precision astrometry allowing for proper motion measurements for relatively nearby and large orbital systems, such as V404 Cygni. By separating the core of the jet from the companion one may be able to independently derive the mass function of compact stellar objects.
- Resolving the spectral jet break. The total luminosity, and hence power of the jet is critically dependent on the exact position of the break in the spectrum. Resolving this region during a state transition will give us a much better handle on how black holes, and other compact objects, power their jets.
- Directly resolving some binary systems: As well as resolving the inner parts of the jets, the extremely high-resolution would resolve the binary companion from the compact object.
- Mapping the structure of the stellar/disk wind: The environments around compact objects are poorly understood and maybe dominated by particle winds from the accretion disk or (if present) a high-mass companion.
- Removing the scattering limit: most XRBs are located along the Galactic plane and many are scatter broadened at longer wavelengths and the inner structure of the source is masked. Therefore, observations at higher wavelength are the only possible method to further resolve some XRBs like Cygnus X-3.

*C44: Pulsars in the Galactic Centre*
M. Spaans[40]

The long term, over years, monitoring of an optimally located pulsar near the Galactic Centre provides a very accurate determination of the central black hole mass M as well as its time derivative dM/dt. If occasional accretion events occur, then dM/dt should show these as spikes (or as steps in mass) and thus be easily identifiable. However, even in the absence of mass accretion, massive black holes can interact with Wheeler's quantum foam of mini black holes at the Planck scale. These vacuum quantum fluctuations constitute a direct expression of the mysterious dark energy that currently seems to drive cosmic expansion (Spaans, M., 2013, Int.J.Mod.Phys.D, Vol.22, No.9, 1330022). Under this quantum interaction, one expects a smooth monotonic increase of Sgr A*'s mass M with time when there is no mass accretion through the event horizon (Spaans, M., 1997, Nuc.Phys.B, 492, 526). This allows one to constrain the properties of the Planck scale vacuum. In general, accurate measurements of M and dM/dt provide tight limits on any quantum thermodynamic behavior of Sgr A*.

*C45: ALMA VLBI with LLAMA[4]*
J. Lepine[64], Z. Abraham[64]

Brazil and Argentine are in the final steps of an agreement to install the 12-m LLAMA radiotelescope in Argentina at a distance of about 150 km from ALMA as part of a VLBI set-up. At higher frequencies mmVLBI experiments with APEX, ASTE, and possibly a millimetric antenna in Peru will be possible. Our scientific interest includes the study of quasars, black holes, interstellar medium, and also the structure of our Galaxy. We would like, in future, to perform parallax measurements of star formation regions in the Galactic arms, similar to those of the VERA project in Japan.

*C46: Probing dark matter sub-halo lensing of galaxies at sub-mm wavelengths*
E. Zackrisson[13], K. Wiik[57], S. Asadi[13], E. Freeland[13]

In the case where a gravitational lens is a galaxy, high-resolution imaging can be used to test for the presence of small-scale structure within its dark matter halo. Simulations suggest that a vast population of extremely faint or completely dark sub-halos in the dwarf- galaxy mass range may be awaiting detection in the vicinity of every large galaxy. Another prospect is to use mmVLBI with ALMA in band 3 to detect halo substructure in the case where the source is an AGN jet. Due to the small intrinsic sizes of such sources, only extremely compact forms of halo substructure (e.g. primordial black holes or ultra- compact mini-halos) predicted by alternative structure formation scenarios can currently be probed. The prospects for detecting standard cold dark matter sub-halos using mmVLBI are likely better when multiple-imaged sub-mm galaxies (with much larger intrinsic sizes) are targeted. Upcoming simulations will be used to explore sub-halo lensing of galaxies at sub-mm wavelengths, where ALMA can be used to improve both the sensitivity and imaging quality of a sub-mm VLBI array.

---
[4] This is a non–European response to our call.

*C47: AGN Jets: spectral shape, turnover frequencies and spectral steepening*
P. Augusto[10], J. Afonso[10]

We are interested in mmVLBI at 43 GHz and higher frequencies to determine the spectral shape, turnover frequencies and measure spectral steepening in AGN. In order to estimate synchrotron ages, magnetic fields and energy densities of components/radio sources we must know where the quasi-linear optically thin component spectrum steepens at high frequencies. This is normally assumed to lie at 100 GHz but (honestly) for the vast majority of the sources this assumption can be wrong! Hence, here resides one of the most crucial roles of a future mmVLBI: confirm/establish the actual standard frequency where the spectrum of powerful radio sources steepens. In reference to future global mmVLBI we note that a suggested new mm-telescope located at Madeira would nicely fill a gap in the uv-plane.

*C48: Nature's coupling constants and CMB temperature as function of redshift*
C. Martins[9], M. Ferreira[9], M. Julião[9], A. Leite[9], A. Monteiro[9], P. Pedrosa[9], M. Silva[9], J. Vieira[9], P. Vielzeuf[9]

Our group is actively involved in constraining the evolution of nature's fundamental couplings throughout the cosmic history (and exploiting its fundamental physics implications), and ALMA measurements using molecular absorption will undoubtedly bring significantly improvements. Equally important in this regard are measurements of the CMB temperature, which sometimes can be made in the same systems. PKS1830-211 is the best example of this. We believe that such measurements will become increasingly important in the coming years. Simultaneously mapping the values of T_CMB and fundamental constants as a function of redshift is hugely important from a fundamental physics point of view and in particular can shed light on the enigma of dark energy. Another area of interest (where we currently have less expertise, but which we plan to strengthen in the coming years) is testing general relativity in the strong field limit). All of the above topics are ideal examples of possible synergy between ALMA and the E-ELT. Tests of GR in the galactic center are a key science driver for ELT-CAM, and fundamental constants and T(z) will be a key science driver for ELT-HIRES.

*C49: ALMA as radar receiver for studying Near Earth objects*
K. van 't Klooster[71]

Recently I proposed to investigate the use of an ESA 35m antenna, like the one in Malargue (Argentina), as powerful transmitter with ALMA as the receiving station for Near Earth Object (NEO) imaging, space surveillance and limited debris capability, with ALMA either
1. As a fully beam-formed reception station, equivalent to ~84 m dish, for NEO and other single targets.
or

2. As a multiple beam receiver station (beam-park) for volume investigations.

This scenario would require a transmission frequency near Ka-band (likely 34 GHz, near the uplink frequency for Deep Space) and ALMA receiving in, likely, band-1 (31-45 GHz). Such configuration could have similar or more powerful capabilities than the 70m/VLA scenario operating at 8.4 GHz (Goldstone–VLA). The project is not a current ESA activity, but a study of its technological feasibility may be (re)submitted for the General Study call for ideas at ESA. With the failure of the high-gain antenna on-board the Galileo mission to Jupiter, NASA requested assistance from VLA and was important in enabling X-band receivers at the VLA. Such situation could also arise at these higher Ka-band frequencies.

# APPENDIX B: Affiliations

1: Abastumani Observatory, Abastumani, Georgia
2: Leiden Observatory, Leiden, The Netherlands
3: Astronomical Institute 'Anton Pannekoek', University of Amsterdam, Amsterdam, The Netherlands
4: Astronomical Observatory, Universitat de València, Valencia, Spain
5: CEA/IRFU/SAp/Saclay, Gif-sur Yvette, France
6: CNRS, LESIA - LUTH, Observatoire de Paris, Université Paris Diderot, Université Pierre et Marie Curie, Paris, France
7: CNRS, Laboratoire d'Astrophysique de Bordeaux, Université de Bordeaux, Floriac, France
8: Centro de Astrobiologia (INTA-CSIC), Torrejon de Ardoz, Spain
9: Centro de Astrofísica da Universidade do Porto, Porto, Portugal
10: Centro de Astronomia e Astrofísica da Universidade de Lisboa, Lisboa, Portugal
11: Departament d'Astronomia i Meteorologia, University de Barcelona, Barcelona, Spain
12: Departamento de Astronomía y Astrofísica, Universitat de València, Valencia, Spain
13: Department of Astronomy, Stockholm University, Stockholm, Sweden
14: Dipartimento di Fisica e Astronomia, Bologna University, Bologna, Italy
15: European Southern Observatory (ESO), Garching, Germany
16: Excellence Cluster Universe, Technische Universität München, Garching, Germany
17: FOMI Satellite Geodetic Observatory, Budapest, Hungary
18: Faculty of Physics, University of Bielefeld, Bielefeld, Germany
19: Frankfurt Institute for Advanced Studies (FIAS), Frankfurt, Germany
20: Hamburger Sternwarte, University of Hamburg, Hamburg, Germany
21: Hvar Observatory, Faculty of Geodesy, University of Zagreb, Croatia
22: IGAM, Institute of Physics, University of Graz, Graz, Austria
23: IMAPP, Radboud University, Nijmegen, The Netherlands
24: INAF - Istituto di Astrofisica Spaziale di Bologna, Italy
25: INAF - Istituto di Astrofisica e Planetologia Spaziali di Roma (IAPS), Roma, Italy
26: INAF - Istituto di Radioastronomia, Bologna, Italy
27: INAF - Osservatorio Astrofisico di Arcetri, Firenze, Italy
28: INAF - Osservatorio Astronomico di Cagliari, Selargius, Italy
29: Institut de Radioastronomie Millimétrique, Grenoble, France
30: Institut für Theoretische Physik und Astrophysik, Würzburg University, Würzburg, Germany
31: Institut für Theoretische Physik, Frankfurt am Main, Germany
32: Institute for Space Sciences, Bucharest, Romania
33: Institute of Astrophysics, Pontificia Universidad Catolica de Chile, Santiago, Chile
34: Institute of Radio Astronomy, National Academy of Sciences of Ukraine, Kharkov, Ukraine
35: Instituto de Astrofísica de Andalucía (IAA-CSIC), Granada, Spain
36: Instituto de Ciencias del Espacio (CSIC)-UB/IEEC, Universitat de Barcelona, Barcelona, Spain
37: Ioffe Physical-Technical Institute, St. Petersburg, Russia
38: JBCA, Dept. Physics and Astronomy, University of Manchester, Manchester, UK
39: Joint Institute for VLBI in Europe, Dwingeloo, The Netherlands
40: Kapteyn Institute, University of Groningen, Groningen, The Netherlands
41: LERMA, Observatoire de Paris, Université Paris-Diderot, Meudon, France
42: MTA Research Centre for Astronomy and Earth Sciences, Konkoly Observatory, Budapest, Hungary
43: Max-Planck-Institut für Extraterrestrische Physik, Garching, Germany
44: Max-Planck-Institut für Radioastronomie, Bonn, Germany
45: Max-Planck-Institute für Gravitational Physics, Golm, Germany
46: Observatoire Astronomique de Strasbourg, Université de Strasbourg, Strasbourg, France
47: Observatorio Astronomico Nacional, E-28014 Madrid, Spain
48: Onsala Space Observatory, Chalmers University of Technology, Onsala, Sweden
49: Osservatorio Astronomico di Trieste, Trieste, Italy
50: Petersburg Nuclear Physics Institute, Gatchina, St. Petersburg, Russia
51: Physikalisches Institut, Universität zu Köln, Köln , Germany

52: Research Institute in Astrophysics and Planetology, University of Toulouse, France
53: Royal Observatory of Edinburgh, Edignburgh, UK
54: School of Physics and Astronomy, Manchester University, UK
55: School of Physics and Astronomy, University of Southampton, Hampshire, UK
56: Sternwarte Bamberg and Erlangen Centre for Astroparticle Physics, University of Erlangen-Nuremberg, Erlangen, Germany
57: Tuorla Observatory, Department of Physics and Astronomy, University of Turku, Piikkiö, Finland
58: University of Hamburg, Hamburg, Germany
59: University of Hertfordshire, Hatfield, UK
60: University of Turin, Turin, Italy
61: Vatican Observatory, Vatican City State, Italy
62: Astronomy Department, Cornell University, Ithaca, NY, USA
63: Departamento de Astronomia, Universidad de Concepción, Biobío, Chile
64: Departamento de Astronomia, University of Sao Paulo, Sao Paulo, Brazil
65: Department of Astronomy, University of Cape Town, Rondebosch, South Africa
66: European Southern Observatory (ESO), Santiago, Chile
67: International Centre for Radio Astronomy Research, Curtin University, Perth, Australia
68: JPL, Caltech, Pasadena, CA, USA
69: NRAO, Charlottesville, VA, USA
70: University of Alberta, Edmonton, Canada
71: ESA-ESTEC, Noordwijk, The Netherlands

52: Research Institute in Astrophysics and Planetology, University of Toulouse, France
53: Royal Observatory of Edinburgh, Edignburgh, UK
54: School of Physics and Astronomy, Manchester University, UK
55: School of Physics and Astronomy, University of Southampton, Hampshire, UK
56: Sternwarte Bamberg and Erlangen Centre for Astroparticle Physics, University of Erlangen-Nuremberg, Erlangen, Germany
57: Tuorla Observatory, Department of Physics and Astronomy, University of Turku, Piikkiö, Finland
58: University of Hamburg, Hamburg, Germany
59: University of Hertfordshire, Hatfield, UK
60: University of Turin, Turin, Italy
61: Vatican Observatory, Vatican City State, Italy
62: Astronomy Department, Cornell University, Ithaca, NY, USA
63: Departamento de Astronomia, Universidad de Concepción, Biobío, Chile
64: Departamento de Astronomia, University of Sao Paulo, Sao Paulo, Brazil
65: Department of Astronomy, University of Cape Town, Rondebosch, South Africa
66: European Southern Observatory (ESO), Santiago, Chile
67: International Centre for Radio Astronomy Research, Curtin University, Perth, Australia
68: JPL, Caltech, Pasadena, CA, USA
69: NRAO, Charlottesville, VA, USA
70: University of Alberta, Edmonton, Canada
71: ESA-ESTEC, Noordwijk, The Netherlands